\documentclass{spie}
\usepackage{amssymb}
\usepackage{amsfonts}
\usepackage{amsmath}
\usepackage{cite}

\input{tcilatex}
\begin{document}

\title{The 3-3-1 model with exotic electric charges, right-handed neutrinos
with type-I+II seesaw mechanism and its effects on LFV}
\author{ Abrar Ahmad\supit{a+}, Shakeel Mahmood\supit{b*}, Farida Tahir%
\supit{a++}, Wasi Uz Zaman\supit{b**} and Fizza Atif\supit{b***}. \\
%EndAName
\supit{a}Department of Physics, COMSATS University Islamabad, Pakistan.\\
\supit{b}Department of Physics, Air University, PAF Complex, E-9, Islamabad,
Pakistan.}
\authorinfo{\\
\supit{+} \textit{E-mail: smart5733@gmail.com,}\\
\supit{*} \textit{E-mail: shakeel.mahmood@mail.au.edu.pk}\\
\supit{++} \textit{E-mail: farida\_tahir@comsats.edu.pk}\\
\supit{**} \textit{E-mail:190198@students.au.edu.pk}\\
\supit{***} \textit{E-mail:fizaatif000@gmail.com}}
\maketitle

\begin{abstract}
In this research, we propose a modified version of the 3-3-1 model,
incorporating a type I+II seesaw mechanism and $Z_{4}$ discrete symmetry, as
a framework for investigating lepton flavor-violating (LFV) decays. This
model successfully yields the left-handed neutrino's mass square difference
in the eV scale, with specific values $\Delta m_{21}^{2}=7.12\times
10^{-5}~eV^{2},$ $\Delta m_{31}^{2}=2.55\times 10^{-3}~eV^{2}$, and
generates mixing angles $\sin ^{2}\theta _{12}=0.304,~\sin ^{2}\theta
_{23}=0.595~$and $\sin ^{2}\theta _{13}=2.15\times 10^{-2}$ that align with
experimental data. Furthermore, we develop SARAH and SPheno \cite{SARAH1}
algorithms tailored for our modified model, enabling us to estimate the
magnitude of LFV observables. Our calculations indicate favorable results
for various LFV branching ratios, including $Br(\mu \rightarrow e\gamma
)=7.82\times 10^{-20},$ $Br(\tau \rightarrow e\gamma )=3.66\times 10^{-22},$ 
$Br(\tau \rightarrow \mu \gamma )=1.03\times 10^{-19},$ $Br(\tau \rightarrow
eee)=3.97\times 10^{-12},$ $Br(\tau \rightarrow \mu \mu \mu )=1.33\times
10^{-12},$ $Br(\tau \rightarrow \mu ee)=8.89\times 10^{-13}$. These findings
demonstrate improved agreement with experimental measurements compared to
previously reported results, which typically fall within the range of $%
10^{-2}$ to $10^{-6}$.
\end{abstract}


\section{Introduction}

Lepton flavor conservation and lepton flavor universality in the standard
model (SM) is not protected by any symmetry, it is just unavailability of
experimental results. Lepton flavor violation (LFV) is predicted in several
physics models \cite{mod1,mod2,mod3,mod5,mod6,mod7}. A finite value of
neutrino mass allows LFV in decays of neutrino-less two body charge leptons
such as $\tau ^{-}\rightarrow l_{i}^{-}\gamma ,~$where $i=1,2$ are denoting
electron and muon respectively. These decays are suppressed due to a factor
of $\frac{m_{\nu }^{2}}{m_{W}^{2}}$ , which produces very small value of $%
10^{-54}$. For neutrino-less decay $\tau \rightarrow
l_{i}^{-}l_{j}^{+}l_{\alpha }^{-},$ where $\alpha =i$ or $j$ and $i$ may or
may not be equal to $j,$ different scenarios have been reported in the
literature contrary with $10^{-54}$ and $10^{-14}$ values respectively \cite%
{res1,res2,res3}. Very high value $10^{-14}$ is discarded due to the revised
calculation by \cite{res6} with a value of $10^{-54}$and \cite{res7} with a
value of $10^{-56}$. Therefore, any observation of charged LFV is reported
new physics (NP) \cite{res5}.

The LFV in the charged sector can shed more light on several unanswered
questions, including that a large baryon number asymmetry, origin of dark
matter and number of fermionic generations are possible. Since LFV in the
charged sector is tied to small neutrino mass. Therefore, understanding the
proper small neutrino mass generation mechanism is important in LFV.

The gauge group of 3-3-1 mode is $SU(3)_{C}\otimes SU(3)_{L}\otimes U(1)_{X}$%
, originally developed in \cite{C1,C2} and it is a straightforward extension
of the SM gauge group. To achieve the symmetry breaking pattern \textit{%
3-3-1 }$\rightarrow SM\rightarrow U(1)_{em}$ further extra scalars are
needed. The first symmetry breaking occurs on a significantly higher scale ($%
\approx $1000 GeV) than that of the EW-scale. The 3-3-1 model has different
variants, and each one can be characterized by a parameter known as $\beta .$
With the different values of $\beta $ along with the anomaly cancellation
condition, do requires new exotic particles having different electric
charges.

\begin{itemize}
\item The 3-3-1 model has anomaly cancellation, together with the asymptotic
freedom of quantum chromodynamics (QCD) and it restricts the ratio of number
of generations to the number of colors, so it gives the existence of three
generations, and the SM is silent about this problem \cite{C3}.

\item Anomaly cancellation do effect the fermionic transformation properties
that is due to the said cancellation, the left-handed fermions transform as
triplets (or antitriplets) by extending $SU(2)_{L}$ to $SU(3)_{L}.$ For
anomaly-free theory, the number of triplets in the fermion sector must be
equal to the number of anti-triplets for nontrivial $SU(3)$ gauge structure.
So, three leptonic generations should transform as an antitriplets, and two
quark's generations transform as triplets and one be an antitriplet (termed
as triplets plus one option). We know that each quark carries three color.
Therefore, the different transformation strength of third generation of
quarks can be a reason for the large top quark mass with respect to other
quarks \cite{C4, Types-of-anomalies1, Types-of-anomalies2}.

\item The extension of $SU(2)_{L}$ to $SU(3)_{L}$ group demands the presence
of five additional gauge bosons. So, a new neutral boson $Z^{^{\prime }}$
and four extra bosons that might be charged or neutral (depending on the
model variant) \cite{C5}.

\item One of the quark generation can transform differently from the other
two to ensure anomaly-free conditions. This lead to tree-level
flavor-changing neutral current (FCNC) interaction via a new neutral gauge
boson $Z^{^{\prime }}$ or the mixing of $Z$ and $Z^{^{\prime }}$ bosons \cite%
{C6}.

\item The coupling constant $g^{\prime }$ and $g_{X}$ of $SU(3)_{L}$ and $%
U(1)_{X}$ respectively are related with electroweak (EW) mixing angle $%
\theta _{W}$ with following ratio;%
\begin{equation}
\frac{g_{X}}{g}=\frac{\sin \theta _{W}}{\sqrt{1-(1+\beta ^{2})\sin
^{2}\theta _{W}}}.  \label{3.1}
\end{equation}%
In above relation, $\beta $ is an arbitrary with coefficient that describes
the nature of model and can be fixed with anomaly-free condition along
fermionic representation. There are four possible values of $\beta =\pm 
\sqrt{3},\pm \frac{1}{\sqrt{3}}$ respectively in literate with different
fermionic representation \cite{C1,C5,arbitrary-beta}. At $\beta =\pm \sqrt{3}%
,$ $\sin ^{2}\theta _{W}=\frac{1}{4}=0.25$ and $\beta =\pm \frac{1}{\sqrt{3}}
$, $\sin ^{2}\theta _{W}=\frac{3}{4}=0.75$, the $U(1)_{X}$ coupling constant 
$g_{X}(\mu )$ diverges (Landau pole appear), and it leads to the breakdown
of the model's perturbative behavior at an energy scale $\mu .$ It imposes a
constraint on mixing angle $\sin ^{2}\theta _{W}<0.25$ and it results in the
complete embedding of $SU(2)_{L}$ as a subgroup within $SU(3)_{L}$ \cite%
{331-model-energy-scale1,331-model-energy-scale2}.

\item The 3-3-1 model has more particles than the SM because of the gauge
group extensions. With five extra gauge bosons, and heavy quarks or leptons,
and six more Higgs scalars, the model gives rich phenomenology at the Large
Hadron Collider (LHC).

\item The version of the 3-3-1 model which predicts doubly charged particles
(leptons, gauge bosons, and Higgs bosons) and new quarks with exotic
electric charges of $\frac{5}{3}e$ and $-\frac{4}{3}e$ \cite{Diphoton-excess}
has right phenomenological aspects.
\end{itemize}

The existing literature on the 3-3-1 model has given various approaches to
generate light neutrino mass \cite{TypeI+II, TypeI+IIb, Mass1, Mass2, Mass3,
Mass4}. Among these, the doubly charged lepton version of the 3-3-1 model
stands out as it accommodates the framework for neutrino mass generation.
This version with is particularly favored because it aligns well with the
potential of Future Circular Collider (FCC)-Based Energy-Frontier \textit{ep}%
-Proton Colliders and Super Proton-Proton Collider (SppC)-based \textit{ep}
colliders, which could provide an excellent environment to search for the
existence of doubly charged leptons in the FCC and SppC \cite{FCC,SppC}.
Therefore \emph{we have mofied the mathematical framework for light
neutrinos mass generation by applying the type-I+II seesaw mechanism along
with }$\emph{Z}_{\emph{4}}$\emph{\ discrete symmetry in doubly charged
lepton version of the 3-3-1 model. Furthermore, we have developed its SARAH
and SPheno }\cite{SARAH1,SARAH2,SARAH3} \emph{algorithm for our modified
3-3-1 model to estimate particles masses and mixing matrices along with LFV
observable by using unique input parameters.}

We have discussed the fermions representation of our model in section 2. The
section 3 is dedicated for the detailed work on spontaneous symmetry
breaking which is necessary for mass generation. In section 4, input
parameters for our calculations are provided. Numerical results obtained by
using Spheno are given in section 5 and finally the conclusion is given in
section 6.

\section{Fermion Representation of Model}

The electric charge operator of 3-3-1 model can be written in the linear
combination of the diagonal generators of $SU(3)_{L}$ and $U(1)_{X}$ group
as:

\begin{equation}
\hat{Q}=\hat{T}^{3}+\beta \hat{T}^{8}+X\hat{I},  \label{3.1.1}
\end{equation}%
where $\hat{I}=Diag(1,1,1)$, $\hat{T}^{i}=\frac{1}{2}\lambda ^{i}$ are the
the generators of $SU(3)_{L}$, here $(\lambda ^{i}$ are Gell-Mann matrices
and $i=1,...8).$ The generators $\hat{T}^{i}$ satisfy the normalization
condition $Tr[\hat{T}^{i},\hat{T}^{j}]=\frac{\delta ^{ij}}{2}.$ For the
anti-triplet $\bar{T}^{^{\ast }i}=-(\hat{T}^{i})^{T}$, $T$ is the transpose
of matrix. The electric charge operator depends upon $\beta $ and $U(1)$
quantum number $X.$\ $\beta $ is not only the arbitrary constant, but also
signifies the nature of the model. In this work $\beta =-\sqrt{3}$ is
considered along with the notations used in \cite%
{Collider-Phenomenology,C1.2}$.$ The fermionic Lagrangian is as follows:%
\begin{equation}
\tciLaplace _{fermion}=i\bar{\psi}^{\alpha }D_{\mu }\gamma ^{\mu }\psi
^{\alpha },  \label{3.3.2}
\end{equation}%
where $\alpha $ represents fermionic flavour and $D_{\mu }$ is the covariant
derivative defined as:%
\begin{equation}
D_{\mu }\equiv \partial _{\mu }-\frac{1}{2}igW_{\mu }^{i}\lambda
^{i}-ig_{X}XX_{\mu }.  \label{3.3.3}
\end{equation}%
The covariant derivative act on the fermionic fields as:

$\rhd $ Triplets $D_{\mu }\psi _{L}:\partial _{\mu }\psi _{L}-\frac{1}{2}%
igW_{\mu }^{i}\lambda ^{i}\psi _{L}-ig_{X}XX_{\mu }\psi _{L},$

$\rhd $ Anti-triplets $D_{\mu }\bar{\psi}_{L}:\partial _{\mu }\bar{\psi}_{L}+%
\frac{1}{2}igW_{\mu }^{i}\left( \lambda ^{i}\right) ^{T}\bar{\psi}%
_{L}-ig_{X}XX_{\mu }\bar{\psi}_{L},$

$\rhd $ Singlets $D_{\mu }\psi _{R}:\partial _{\mu }\psi _{R}-ig_{X}XX_{\mu
}\psi _{R}.$

Here $W_{\mu }^{i}$ and $X_{\mu }$ represent gauge fields of $SU(3)_{L}$ and 
$U(1)_{X}$ respectively. For all the left-handed (LH) components of three
generations leptons fields transformed as anti-triplet whereas their
corresponding right-handed (RH) components transformed as singlets. In the
case of quarks fields, the two initial generations of LH components of
quarks fields transformed as triplets, LH components of third generation of
quarks field transformed as anti-triplet and RH components of quarks fields
transformed as singlets respectively.

Fermionic fields with their respective $SU(3)_{C},$ $SU(3)_{L}$ and $%
U(1)_{X} $ quantum numbers respectively are represented as:

\emph{Lepton:}%
\begin{eqnarray}
l_{_{\alpha }L} &=&\left( 
\begin{array}{c}
e_{\alpha } \\ 
-\nu _{\alpha } \\ 
E_{\alpha }%
\end{array}%
\right) _{L}\sim \left( 1,3^{\ast },-1\right) ,\text{ }\alpha =1,2,3,  \notag
\\
e_{\alpha R} &=&\left( 1,1,-1\right) ,\text{ }\nu _{\alpha R}=\left(
1,1,0\right) ,\text{ }E_{\alpha R}=\left( 1,1,-2\right) ,  \label{3.3.4}
\end{eqnarray}%
$E_{\alpha L,R}$ heavy exotic leptons with $-2e$ electric charge and three
RH neutrinos $\nu _{\alpha R}$ are introduced in this version of 3-3-1 model.

\emph{Quarks:} 
\begin{eqnarray}
Q_{iL} &=&\left( 
\begin{array}{c}
u_{i} \\ 
d_{i} \\ 
J_{i}%
\end{array}%
\right) _{L}\sim \left( 3,3,\frac{2}{3}\right) ,\text{ }i=1,2,  \notag \\
u_{iR} &=&\left( 3,1,\frac{2}{3}\right) ,\text{ }d_{iR}=\left( 3,1,-\frac{1}{%
3}\right) ,\text{ }J_{iR}=\left( 3,1,\frac{5}{3}\right) ,  \label{3.3.5}
\end{eqnarray}%
\begin{eqnarray}
Q_{3L} &=&\left( 
\begin{array}{c}
d_{3} \\ 
-u_{3} \\ 
J_{3}%
\end{array}%
\right) _{L}\sim \left( 3,3^{\ast },-\frac{1}{3}\right) ,\text{ }  \notag \\
d_{3R} &=&\left( 3,1,-\frac{1}{3}\right) ,\text{ }u_{3R}=\left( 3,1,\frac{2}{%
3}\right) ,\text{ }J_{3R}=\left( 3,1,-\frac{4}{3}\right) ,  \label{3.3.6}
\end{eqnarray}%
where $J_{i}$ and $J_{3}$ are heavy exotic quarks with electric charge $%
\frac{5}{3}e$ and $-\frac{4}{3}e$ respectively as predicted by model.

\subsection{Fermionic anomalies cancellation}

The term "anomaly" in literature is defined as:

\begin{itemize}
\item[a)] "Anomaly (in something) refers to a thing, situation, etc. that is
different from what is normal or expected"\cite{def1}.

\item[b)] "Deviation from the common rule"\cite{def2}.
\end{itemize}

So, anomaly is a deviation from established normal pattern. Spotting the
anomaly depends on the ability to define what is normal.

An anomaly occurs when quantum corrections do not respect a symmetry of the
classical Lagrangian.

In other words, the symmetry present in the classical Lagrangian is broken
by quantum effects. This occurs if the action of a theory is invariant under
a symmetry. Anomalies are an important tool to identify the existence of new
proposals for physics beyond SM.

In 4-D space-time, following three perturbative anomalies associated with
chiral gauge theories must be eliminated \cite%
{Types-of-anomalies1,Types-of-anomalies2}.

\begin{enumerate}
\item[i)] "The Triangular Chiral Gauge Anomaly" (To prevent gauge invariance
and renormalizability problems).

\item[ii)] "The Global Non perturbative $SU(2)$ Chiral Gauge Anomaly" (In
order for the fermion integral to be well defined).

\item[iii)] "The Mixed Perturbative Chiral Gauge Gravitational Anomaly" (In
order to assure generic covariance).
\end{enumerate}

In order to cancel anomalies we must have to sum over all the hypercharges
of quarks and leptons. Here one question arises why we have to sum over all
hypercharges? Let us see in SM, since the weak isospin group $SU(2)$ is a
special unitary group so group demands two things that it is unitary group
of rank two and having determinant unitary. So, for the unitary determinant
we have following trace;%
\begin{equation}
Tr[\{ \hat{T}^{i},\hat{T}^{j}\} \hat{T}^{k}]=2\delta ^{ij}Tr(\hat{T}^{k})=0,
\label{D6}
\end{equation}%
where $\hat{T}^{i},~\hat{T}^{j}$ and $\hat{T}^{k}$ are the three generators
of group. Here at least one generator is hypercharge $Y$ so above equation
transform as:%
\begin{equation}
Tr[\{ \hat{T}^{i},\hat{T}^{j}\}Y]=2\delta ^{ij}Tr(Y)=0.  \label{D7}
\end{equation}%
As trace should be zero for $SU(2)$ representation and we can only see
electromagnetic charge, so, we sum over all the hypercharges. On enlarging
gauge group, the fermionic multiplets representation mentioned in Eqs. (\ref%
{3.3.4} and \ref{3.3.5}) satisfy the gauge anomaly free constraint equations
provided by vertices $[SU(3)_{C}]^{3},~[SU(3)_{L}]^{3}$ $%
[SU(3)_{C}]^{2}~U(1)_{X}$,$~[SU(3)_{L}]^{2}~U(1)_{X},$ $U(1)_{X}^{3},$ $%
[gravitation]^{2}U(1)_{X}$ \cite{C3,generalized-331-models,arbitrary-beta}.
We will check anomaly status of each vertex as follows.

\begin{itemize}
\item $[SU(3)_{C}]^{3}:$\textbf{\ Demands that number LH
triplets(anti-triplets) of color should be equal to number of RH triplet.}%
\begin{eqnarray}
3(\text{No.~of }Q_{triplet}+\text{No.~of}~Q_{anti-triplet})-\text{No.~of }Q_{%
\func{Si}nglet}. &=&0,  \label{D8} \\
3\left( \tsum \limits_{i=1}^{2}(Q_{iL}+Q_{3L}\right) -\left( \text{ }\tsum
\limits_{\alpha =1}^{3}u_{\alpha R}+\tsum \limits_{\alpha =1}^{3}d_{\alpha
R}+\tsum \limits_{i=1}^{2}j_{iR}+j_{3R}\right) &=&0,  \notag
\end{eqnarray}%
\begin{equation*}
3(2+1)-(3+3+3)=0.
\end{equation*}

\item $[SU(3)_{L}]^{3}:$\textbf{\ Demands that number triplets should be
equal to number of anti-triplet.}%
\begin{equation}
\text{No.~of }l_{triplet}-\text{No.~of}~l_{anti-triplet}+3(\text{No.~of }%
Q_{triplet}-\text{No.~of}~Q_{anti-triplet})=0  \label{D9}
\end{equation}%
\begin{equation*}
0-3+3(2-1)=0
\end{equation*}

\item $[SU(3)_{C}]^{2}[U(1)_{X}]:$\textbf{\ Demands that trace of }$U(1)_{X}$%
\textbf{\ charge over the color fermions should be zero.}%
\begin{equation}
3\tsum \limits_{\text{generation}}X_{Q_{L}}-\tsum \limits_{\sin \text{glet}%
}X_{q_{R}}=0,  \label{D10}
\end{equation}%
\begin{eqnarray*}
3\left( \tsum \limits_{i=1}^{2}X_{Q_{iL}}+X_{Q_{3L}}\right) -\tsum
\limits_{\alpha =1}^{3}\left( X_{u_{\alpha R}}-X_{d_{\alpha R}}\right)
-\tsum \limits_{i=1}^{2}X_{j_{iR}}-X_{J_{3R}} &=&0, \\
3\left[ 2\left( \frac{2}{3}\right) +\left( -\frac{1}{3}\right) \right]
-3\left( \frac{2}{3}\right) -3\left( -\frac{1}{3}\right) -2\left( \frac{5}{3}%
\right) -\left( -\frac{4}{3}\right) &=&0.
\end{eqnarray*}

\item $[SU(3)_{L}]^{2}[U(1)_{X}]:$\textbf{\ Demands that trace of }$U(1)_{X}$%
\textbf{\ charge over the left LH fermions should be zero,}%
\begin{eqnarray}
3\tsum \limits_{\text{generation}}X_{Q_{L}}+\tsum \limits_{\text{generation}%
}X_{l_{L}} &=&0,  \label{D11} \\
3\left( \tsum \limits_{i=1}^{2}X_{Q_{iL}}+X_{Q_{3L}}\right) +\tsum
\limits_{\alpha =1}^{3}X_{l_{\alpha L}} &=&0,  \notag \\
3\left[ 2\left( \frac{2}{3}\right) +\left( -\frac{1}{3}\right) \right]
+3(-1) &=&0.  \notag
\end{eqnarray}

\item $[U(1)_{X}]^{3}:$\textbf{\ Demands that trace of }$[U(1)_{X}]^{3}$%
\textbf{\ charge over the all fermions should be zero,}%
\begin{equation}
9\tsum \limits_{\text{generation}}X_{Q_{L}}^{3}-3\tsum \limits_{\text{%
generation}}X_{q_{R}}^{3}+3\tsum \limits_{\text{generation}%
}X_{l_{L}}^{3}-\tsum \limits_{\text{generation}}X_{l_{R}}^{3}=0,  \label{D12}
\end{equation}%
\begin{equation*}
\begin{array}{c}
9\left( \tsum \limits_{i=1}^{2}X_{Q_{iL}}^{3}+X_{Q_{3L}}^{3}\right) -3\left(
\tsum \limits_{\alpha =1}^{3}\left( X_{u_{\alpha R}}^{3}+X_{d_{\alpha
R}}^{3}\right) +\tsum \limits_{i=1}^{2}X_{j_{iR}}^{3}+X_{J_{3R}}^{3}\right)
\\ 
+3\tsum \limits_{\alpha =1}^{3}X_{l_{\alpha L}}^{3}-\tsum \limits_{\alpha
=1}^{3}\left( X_{e_{\alpha R}}^{3}+X_{E_{\alpha R}}^{3}+X_{\nu _{\alpha
R}}^{3}\right) =0,%
\end{array}%
\end{equation*}%
\begin{equation*}
\begin{array}{c}
9\left[ 2\left( \frac{2}{3}\right) ^{3}+\left( -\frac{1}{3}\right) ^{3}%
\right] -3\left[ 3\left( \frac{2}{3}\right) ^{3}+3\left( -\frac{1}{3}\right)
^{3}+2\left( \frac{5}{3}\right) ^{3}+\left( -\frac{4}{3}\right) ^{3}\right]
\\ 
+9(-1)^{3}-3\left[ (-1)^{3}+(-2)^{3}\right] =0.%
\end{array}%
\end{equation*}

\item $[Gravitation]^{2}[U(1)_{X}]:$\textbf{\ Demands that trace of }$%
U(1)_{X}$\textbf{\ charge over the all fermions should be zero,}%
\begin{equation}
9\tsum \limits_{\text{generation}}X_{Q_{L}}-3\tsum \limits_{\text{generation}%
}X_{q_{R}}+3\tsum \limits_{\text{generation}}X_{l_{L}}-\tsum \limits_{\text{%
generation}}X_{l_{R}}=0,  \label{D13}
\end{equation}%
\begin{eqnarray*}
&&%
\begin{array}{c}
9\left( \tsum \limits_{i=1}^{2}X_{Q_{iL}}+X_{Q_{3L}}\right) -3\left( \tsum
\limits_{\alpha =1}^{3}\left( X_{u_{\alpha R}}+X_{d_{\alpha R}}\right)
+\tsum \limits_{i=1}^{2}X_{j_{iR}}+X_{J_{3R}}\right) \\ 
+3\tsum \limits_{\alpha =1}^{3}X_{l_{\alpha L}}-\tsum \limits_{\alpha
=1}^{3}\left( X_{e_{\alpha R}}+X_{E_{\alpha R}}+X_{\nu _{\alpha R}}\right)
=0,%
\end{array}
\\
&&%
\begin{array}{c}
9\left[ 2\left( \frac{2}{3}\right) +\left( -\frac{1}{3}\right) \right] -3%
\left[ 3\left( \frac{2}{3}\right) +3\left( -\frac{1}{3}\right) +2\left( 
\frac{5}{3}\right) +\left( -\frac{4}{3}\right) \right] \\ 
+9(-1)-3\left[ (-1)+(-2)\right] =0.%
\end{array}%
\end{eqnarray*}
\end{itemize}

So fermionic representation mentioned in Eq.(\ref{3.3.4} and \ref{3.3.5}) is
anomaly free.

\section{Spontaneous Symmetry Breaking}

\subsection{Scalar Content}

In the 3-3-1 model, the scalar sector should be considered in addition to
the cancellation of anomalies. The Yukawa Lagrangian connects this sector to
the fermions and gives extra limitations to the quantum numbers. The scalar
fields required to provide fermions with masses are chosen based on the
Yukawa Lagrangian. One key condition for the 3-3-1 model is that it should
meet with SM at EW scale. This demonstrates that at a larger energy scale
spontaneously symmetry breaking of $SU(3)_{L}$ to $SU(2)_{L}$ is acceptable.
In this case spontaneous symmetry breaking (SSB) can be achieved in two
stages as \cite{scalar-sector}:%
\begin{equation*}
SU(3)_{C}\otimes SU(3)_{L}\otimes U(1)_{X}\text{ }\underrightarrow{\chi }%
\text{ }SU(3)_{C}\otimes SU(s)_{L}\otimes U(1)_{Y}\underrightarrow{\text{ }%
\Phi _{SM}}\text{ }SU(3)_{C}\otimes U(1)_{Q}.
\end{equation*}%
In the $1^{st}$ stage vacuum expectation value (VEV) of $\chi $ break
symmetry spontaneously and satisfied the following conditions \cite%
{C5,scalar-sector};%
\begin{equation}
\left[ \hat{T}^{1},\hat{T}^{2},\hat{T}^{3},(\beta \hat{T}^{8}+X\hat{I}%
)\equiv \frac{\hat{Y}}{2}\right] \left \langle \chi \right \rangle =0,
\label{S1a}
\end{equation}%
\begin{equation}
\left[ \hat{T}^{4},\hat{T}^{5},\hat{T}^{6},\hat{T}^{7},(\beta \hat{T}^{8}-X%
\hat{I})\right] \left \langle \chi \right \rangle \neq 0.  \label{S1b}
\end{equation}%
The weak hypercharge of SM is represented by $\hat{Y}$. Equations (\ref{S1a}%
, \ref{S1b}) reflect that 5 gauge bosons gain masses and in order to
maintain the number of degrees of freedom, at least five Goldstone Bosons
are required. The scalar content should be able to produce new fermion
masses through Yukawa Couplings in the $1^{st}$ stage of SSB. The Yukawa
terms should be invariant (1, 1, 0) under the said gauge group, which
restricts the possible representations of the scalar fields%
\begin{eqnarray}
\bar{Q}_{iL}J_{iR}\chi &=&(1,1,0),  \notag \\
\left( 3^{\ast },3^{\ast },-\frac{2}{3}\right) \left( 3,1,\frac{5}{3}\right)
\chi &=&(1,1,0),  \notag \\
\left( 1,3^{\ast },1\right) \chi &=&(1,1,0),  \notag \\
implies\text{ }\chi &\rightarrow &\left( 1,3,-1\right) .  \label{S0}
\end{eqnarray}%
By using Eqs. (\ref{3.1.1}, \ref{S1a}, \ref{S1b} and \ref{S0}) charges and
VEV of scalar $\chi $ defined as%
\begin{equation}
\chi =\left( 
\begin{array}{c}
\chi ^{-} \\ 
\chi ^{--} \\ 
\chi ^{0}%
\end{array}%
\right) \sim \left( 1,3,-1\right) ,\left \langle \chi \right \rangle =\frac{1%
}{\sqrt{2}}\left( 
\begin{array}{c}
0 \\ 
0 \\ 
\upsilon _{\chi }%
\end{array}%
\right) .  \label{S1}
\end{equation}%
In the $2^{nd}$ stage VEV of $\Phi _{SM}$ break symmetry spontaneously and
satisfied the following conditions \cite{C5,scalar-sector};%
\begin{eqnarray}
\left[ \hat{T}^{1},\hat{T}^{2},\left( \hat{T}^{3}-\frac{\hat{Y}}{2}\right) %
\right] \left \langle \Phi _{SM}\right \rangle &\neq &0\text{,}  \label{S2a}
\\
\left[ \hat{Q}\equiv \hat{T}^{3}+\frac{\hat{Y}}{2}\right] \left \langle \Phi
_{SM}\right \rangle &=&0\text{.}  \label{S2b}
\end{eqnarray}%
In the $2^{nd}$ stage of SSB, the scalar content must be able to generate SM
fermion's masses through the Yukawa couplings. The gauge invariant Yukawa
terms restrict the possible representations of the scalar fields needed in
the $2^{nd}$ stage.%
\begin{eqnarray}
\bar{Q}_{iL}u_{iR}\Phi _{SM} &=&(1,1,0)=\bar{Q}_{iL}d_{iR}\Phi _{SM},  \notag
\\
\left( 3^{\ast },3^{\ast },-\frac{2}{3}\right) \left( 3,1,\frac{2}{3}\right)
\Phi _{SM} &=&(1,1,0)=\left( 3^{\ast },3^{\ast },-\frac{2}{3}\right) \left(
3,1,-\frac{1}{3}\right) \Phi _{SM},  \notag \\
\left( 1,3^{\ast },0\right) \Phi _{SM} &=&(1,1,0)=\left( 1,3^{\ast
},-1\right) \Phi _{SM},  \notag \\
&\Rightarrow &\text{ }\Phi _{SM}\rightarrow \left( 1,3,0\right) \text{ and }%
\Phi _{SM}\rightarrow \left( 1,3,1\right) .  \label{S2}
\end{eqnarray}%
Eq. (\ref{S2}) show that we need two scalar triplet in the $2^{nd}$ stage of
SSB labeled as $\eta =\left( 1,3,0\right) $ and $\rho =\left( 1,3,1\right) .$
By using Eqs. (\ref{3.1.1}, \ref{S2a}, \ref{S2b} and \ref{S2}) the VEV and
charge content of scalar $\eta $ and $\rho $ defined as%
\begin{eqnarray}
\eta &=&\left( 
\begin{array}{c}
\eta ^{0} \\ 
\eta ^{-} \\ 
\eta ^{+}%
\end{array}%
\right) \sim \left( 1,3,0\right) ,\text{ \ }\left \langle \eta \right
\rangle =\frac{1}{\sqrt{2}}\left( 
\begin{array}{c}
\upsilon _{\eta } \\ 
0 \\ 
0%
\end{array}%
\right) ,  \label{S3} \\
\rho &=&\left( 
\begin{array}{c}
\rho ^{+} \\ 
\rho ^{0} \\ 
\rho ^{++}%
\end{array}%
\right) \sim \left( 1,3,1\right) ,\text{ \ }\left \langle \rho \right
\rangle =\frac{1}{\sqrt{2}}\left( 
\begin{array}{c}
0 \\ 
\upsilon _{\rho } \\ 
0%
\end{array}%
\right) .  \label{S4}
\end{eqnarray}

\subsection{Gauge Sector}

The gauge sector Lagrangian is \cite{Collider-Phenomenology}%
\begin{equation}
\tciLaplace _{gauge}=-\frac{1}{4}F_{\mu \nu }^{a}F^{a\mu \nu }-\frac{1}{4}%
B_{\mu \nu }B^{\mu \nu },  \label{G1}
\end{equation}%
where 
\begin{equation*}
F_{\mu \nu }^{i}=\partial _{\mu }W_{\nu }^{i}-\partial _{\nu }W_{\mu
}^{i}+g^{\prime }f_{ijk}W_{\mu }^{j}W_{\nu }^{k},
\end{equation*}%
and 
\begin{equation*}
B_{\mu \nu }=\partial _{\mu }X_{\nu }-\partial _{\nu }X_{\mu },
\end{equation*}%
are field strength tensor of $SU(3)_{L}$ and $U(1)_{X}$ group respectively.
The antisymmetric structure constants $f_{ijk}$ calculated by using the
algebra%
\begin{equation}
f_{ijk}=\frac{1}{4i}Tr\left[ \left \{ \lambda _{i},\lambda _{j}\right \}
\lambda _{k}\right] .  \label{G4}
\end{equation}%
Gauge bosons get their masses from the kinetic terms of the scalar fields
after SSB%
\begin{equation}
\tciLaplace _{scalar}^{kinetic}=\left( D_{\mu }\chi \right) ^{\dag }\left(
D^{\mu }\chi \right) +\left( D_{\mu }\rho \right) ^{\dag }\left( D^{\mu
}\rho \right) +\left( D_{\mu }\eta \right) ^{\dag }\left( D^{\mu }\eta
\right) ,  \label{G5}
\end{equation}%
where the covariant derivative $D_{\mu }$ is defined in Eq. (\ref{3.3.3}).
The $W_{\mu }^{i}\lambda ^{i}$ matrix in Eq. (\ref{3.3.3}) for triplet
representation is,%
\begin{equation}
W_{\mu }^{i}\lambda ^{i}=\left( 
\begin{array}{ccc}
W_{\mu }^{3}+\frac{1}{\sqrt{3}}W_{\mu }^{8} & -\sqrt{2}W_{\mu }^{+} & -\sqrt{%
2}V_{\mu }^{-} \\ 
-\sqrt{2}W_{\mu }^{-} & -W_{\mu }^{3}+\frac{1}{\sqrt{3}}W_{\mu }^{8} & -%
\sqrt{2}U_{\mu }^{--} \\ 
-\sqrt{2}V_{\mu }^{+} & -\sqrt{2}U_{\mu }^{++} & \frac{-2}{\sqrt{3}}W_{\mu
}^{8}%
\end{array}%
\right) ,  \label{G6}
\end{equation}%
where the charged gauge boson's mass eigenstates are;%
\begin{equation}
W_{\mu }^{\mp }=-\frac{\left( W_{\mu }^{1}\pm iW_{\mu }^{2}\right) }{\sqrt{2}%
},\text{ }V_{\mu }^{\pm }=\frac{-\left( W_{\mu }^{4}\pm iW_{\mu }^{5}\right) 
}{\sqrt{2}},\text{ }U_{\mu }^{\pm \pm }=\frac{-\left( W_{\mu }^{6}\pm
iW_{\mu }^{7}\right) }{\sqrt{2}}.  \label{G7}
\end{equation}%
The scalar field $\chi ~$gains VEV after $1^{st}$ stage of SSB and mass
terms appear from $\left( D_{\mu }\chi \right) ^{\dag }\left( D^{\mu }\chi
\right) $ part of $\tciLaplace _{scalar}^{kinetic}$ as:%
\begin{eqnarray}
m_{V^{\pm }}^{2} &=&\frac{1}{4}g^{2}\upsilon _{\chi }^{2},\text{ \  \ }%
m_{U^{\pm \pm }}^{2}=\frac{1}{4}g^{2}\upsilon _{\chi }^{2},\text{ \  \  \ }%
m_{Z^{^{\prime }}}^{2}=\left( \frac{g^{2}}{3}+g_{X}^{2}\right) \upsilon
_{\chi }^{2},  \label{G8} \\
m_{W^{\pm }}^{2} &=&0,\text{ \ }m_{W^{3}}^{2}=0,\text{ \ }m_{B}^{2}=0.
\label{Ga}
\end{eqnarray}%
Two neutral gauge bosons $B_{\mu }$ and $Z_{\mu }^{\prime }$ are produced by
the mixing of $W_{\mu }^{8}$ and $X_{\mu }$ bosons \cite%
{Collider-Phenomenology}. The mixing is given as: 
\begin{equation}
\binom{B_{\mu }}{Z_{\mu }^{\prime }}=\left( 
\begin{array}{cc}
\sin \theta _{331} & \cos \theta _{331} \\ 
\cos \theta _{331} & -\sin \theta _{331}%
\end{array}%
\right) \binom{X_{\mu }}{W_{\mu }^{8}},  \label{G9}
\end{equation}%
where 
\begin{equation}
\sin \theta _{331}=\frac{g}{\sqrt{g^{2}+3g_{X}^{2}}},~\cos \theta _{331}=%
\frac{\sqrt{3}g_{X}}{\sqrt{g^{2}+3g_{X}^{2}}}.  \label{G10}
\end{equation}%
The scalar field $\eta $ and $\rho $ gains VEV after $2^{nd}$ stage of SSB
hence introduce additional mass terms from $\left( D_{\mu }\eta \right)
^{\dag }\left( D^{\mu }\eta \right) $ and $\left( D_{\mu }\rho \right)
^{\dag }\left( D^{\mu }\rho \right) $ part of $\tciLaplace
_{scalar}^{kinetic}.$ The masses of the charged gauge \cite%
{Collider-Phenomenology} bosons become%
\begin{eqnarray}
m_{W^{\pm }}^{2} &=&\frac{1}{4}g^{2}\left( \upsilon _{\eta }^{2}+\upsilon
_{\rho }^{2}\right) ,  \label{G10a} \\
m_{V^{\pm }}^{2} &=&\frac{1}{4}g^{2}\left( \upsilon _{\chi }^{2}+\upsilon
_{\eta }^{2}\right) ,  \label{G10b} \\
m_{U^{\pm \pm }}^{2} &=&\frac{1}{4}g^{2}\left( \upsilon _{\chi
}^{2}+\upsilon _{\rho }^{2}\right) .  \label{G10c}
\end{eqnarray}%
The consistency condition with SM gives \cite{C1.2} 
\begin{equation}
\upsilon _{\eta }^{2}+\upsilon _{\rho }^{2}\equiv \upsilon ^{2}\simeq
246GeV^{2}.  \label{G11a}
\end{equation}%
The charged boson $W^{\pm }$ should be recognized with the SM charged gauge
boson since it has a mass of the order of the electroweak symmetry breaking
scale, whereas the new bosons are heavier.

In $2^{nd}$ stage, the mixing of $B_{\mu }$ and $W_{\mu }^{3}$ produces a
massless photon $A_{\mu }$ and the SM neutral gauge gauge boson $Z_{\mu }$
along with new constants;%
\begin{equation}
g_{Y}=\frac{g}{\sqrt{3}}\cos \theta _{331}=\frac{gg_{X}}{\sqrt{%
g^{2}+3g_{X}^{2}}},  \label{G11}
\end{equation}%
and the mixing matrix, 
\begin{equation}
\left( 
\begin{array}{c}
Z_{\mu } \\ 
A_{\mu } \\ 
Z_{\mu }^{\prime }%
\end{array}%
\right) =\left( 
\begin{array}{ccc}
\cos \theta _{W} & \sin \theta _{W} & 0 \\ 
-\sin \theta _{W} & \cos \theta _{W} & 0 \\ 
0 & 0 & 1%
\end{array}%
\right) \left( 
\begin{array}{c}
W_{\mu }^{3} \\ 
B_{\mu } \\ 
Z_{\mu }^{\prime }%
\end{array}%
\right) ,  \label{G12}
\end{equation}%
where the Weinberg mixing angle is defined as:%
\begin{equation}
\sin \theta _{W}=\frac{g_{Y}}{\sqrt{g^{2}+g_{Y}^{2}}},\text{ \ }\cos \theta
_{W}=\frac{g}{\sqrt{g^{2}+g_{Y}^{2}}}.  \label{G13}
\end{equation}%
The two physical neutral states $Z_{1\mu }$ and $Z_{2\mu }$ are mixed state
of the SM and heavy gauge bosons $Z_{\mu }$ and $Z_{\mu }^{\prime }.~$The
mass eigentstates of these neutral gauge bosons are obtained with a
additional rotation: 
\begin{equation}
\sin \theta _{z-z^{\prime }}=\frac{\sqrt{3(g^{2}+g_{Y}^{2})(g^{2}-3g_{Y}^{2})%
}}{4g^{2}}\left( \frac{\upsilon _{\eta }^{2}-\upsilon _{\rho }^{2}}{\upsilon
_{\chi }^{2}}-\frac{3g_{Y}^{2}}{g^{2}}\frac{\left( \upsilon _{\eta
}^{2}+\upsilon _{\rho }^{2}\right) }{\upsilon _{\chi }^{2}}\right) ,
\label{G14}
\end{equation}%
that gives%
\begin{equation}
\left( 
\begin{array}{c}
A_{\mu } \\ 
Z_{\mu }^{1} \\ 
Z_{\mu }^{2}%
\end{array}%
\right) =\left( 
\begin{array}{ccc}
1 & 0 & 0 \\ 
0 & \cos \theta _{z-z^{\prime }} & \sin \theta _{z-z^{\prime }} \\ 
0 & -\sin \theta _{z-z^{\prime }} & \cos \theta _{z-z^{\prime }}%
\end{array}%
\right) \left( 
\begin{array}{c}
A_{\mu } \\ 
Z_{\mu } \\ 
Z_{\mu }^{\prime }%
\end{array}%
\right) .  \label{G15}
\end{equation}%
Here it is worth to mention that the phenomenological effect of $Z-Z^{\prime
}$ mixing is not important because the mixing angle $\theta _{z-z^{\prime
}}\lesssim O(10^{-3})$ \cite{ZZ-mixing}, and hence one can approximate $%
Z_{\mu }\approx Z_{\mu }^{1}$ and $Z_{\mu }^{\prime }\approx Z_{\mu }^{2},$
getting the following masses \cite{Collider-Phenomenology}:%
\begin{eqnarray}
m_{A}^{2} &=&0,  \label{G16} \\
m_{Z}^{2} &=&\frac{1}{4\cos ^{2}\theta _{W}}g^{2}\left( \upsilon _{\eta
}^{2}+\upsilon _{\rho }^{2}\right) ,  \label{G17} \\
m_{Z^{\prime }}^{2} &=&\frac{g^{2}\cos ^{2}\theta _{W}\upsilon _{\chi }^{2}}{%
3\left( 1-4\sin ^{2}\theta _{W}\right) }\left[ 
\begin{array}{c}
1+\frac{\left( \upsilon _{\eta }^{2}+\upsilon _{\rho }^{2}\right) }{%
4\upsilon _{\chi }^{2}}\left( 1+\frac{9\sin ^{4}\theta _{W}}{\cos ^{4}\theta
_{W}}\right) \\ 
+\frac{\left( \upsilon _{\eta }^{2}-\upsilon _{\rho }^{2}\right) }{4\upsilon
_{\chi }^{2}}\frac{3\sin ^{2}\theta _{W}}{\cos ^{2}\theta _{W}}%
\end{array}%
\right] .  \label{G18}
\end{eqnarray}%
The mass of the new boson $Z_{0}^{\prime }$ is at the scale $\upsilon _{\chi
}$, which is substantially larger than the SM masses scale. In compact form
the symmetry breaking $SU(3)_{L}\otimes U(1)_{X}$ $\overset{\upsilon _{\chi }%
}{\longrightarrow }$ $SU(2)_{L}\otimes U(1)_{Y}\overset{\upsilon _{\eta
},\upsilon _{\rho }}{\longrightarrow }$ $U(1)_{Q}$ corresponding the
following transformation of the neutral gauge bosons, from the original
basis to the final physical one \cite{C1.2}: $X_{\mu },$ $W_{\mu }^{3},$ $%
W_{\mu }^{8}~\overset{\theta _{331}}{\longrightarrow }\ B_{\mu },$ $W_{\mu
}^{3},$ $Z_{\mu }^{\prime }~\overset{\theta _{W}}{\longrightarrow }~\ A_{\mu
},$ $Z_{\mu },$ $Z_{\mu }^{\prime }~\overset{\theta _{z-z^{\prime }}}{%
\longrightarrow }\ A_{\mu },$ $Z_{\mu }^{1},$ $Z_{\mu }^{2}.$

\subsection{Yukawa Interactions and $Z_{4}$ Discrete Symmetry}

A very interesting relation between $g_{X}$ and $g$ can be derived from Eqs.
(\ref{G11} and \ref{G13}) as;%
\begin{equation}
\frac{g_{X}}{g}=\frac{\sin \theta _{W}}{\sqrt{1-4\sin ^{2}\theta _{W}}}.
\label{Y1}
\end{equation}%
This relationship demonstrates that the theory has a Landau pole. Whereas,
the restriction on the mixing angle $\sin ^{2}\theta _{W}<0.25$, is obtained
to maintain the theory within the perturbative regime. Further, it is
translated into terms of an energy scale of 4-5TeV \cite%
{331-model-energy-scale1,331-model-energy-scale2}. Consequently the
perturbative zone of the model lasts to a few TeV's. In other words, the
model can be considered an effective model up to an energy scale of around
4-5 TeV. Above this threshold, the underlying basic theory must be
discussed. This is the largest energy scale at which the model is
perturbatively trustworthy. It is noteworthy for neutrino masses, because
after the perturbative scale reaches about 4-5 TeV, the effective dimension
five operator $\frac{Y}{\Lambda }\left( \bar{f}_{L}^{C}\rho ^{\ast }\right)
\left( \rho ^{\dag }f_{L}\right) $ \cite{neutrino-mass}, produces the
neutrinos mass formula $m_{\nu }=\frac{Y^{\nu }\upsilon _{\eta }^{2}}{%
\Lambda }.$ For $\upsilon _{\rho }^{2}\approx 100GeV,$ $\Lambda =5TeV,$ we
get neutrino's mass $10Y^{\nu }$ $GeV.$ In other words, heavy neutrino's
mass is produced by the dimension-five operator. To avoid the unpleasant
effect of dimension five operator, we impose $Z_{4}$ discrete symmetry with
the following field transformations:%
\begin{eqnarray}
Q_{iL} &\rightarrow &\omega Q_{iL},~\  \  \ Q_{3L}\rightarrow \omega
^{4}Q_{3L},~\  \  \ l_{_{\alpha }L}\rightarrow \omega l_{_{\alpha }L},  \notag
\\
u_{\alpha R} &\rightarrow &\omega ^{3}u_{\alpha R},~\  \  \ d_{\alpha
R}\rightarrow \omega ^{2}d_{\alpha R},~\  \ J_{iR}\rightarrow \omega
^{2}J_{iR},~\  \  \ J_{3R}\rightarrow \omega ^{3}J_{3R},  \notag \\
e_{\alpha R} &\rightarrow &\omega ^{3}e_{\alpha R},\text{ \  \ }E_{\alpha
R}\rightarrow \omega ^{4}E_{\alpha R},~\  \  \  \nu _{\alpha R}\rightarrow
\omega ^{4}\nu _{\alpha R},  \notag \\
\rho &\rightarrow &\omega ^{3}\rho ,\text{ \  \ }\eta \rightarrow \omega
^{2}\eta ,\text{ \  \ }\chi \rightarrow \omega ^{3}\chi ,  \label{Y2}
\end{eqnarray}%
where $\alpha =1,2,3.$ $i=1,2$ $\ $and $\omega =e^{i\frac{\pi }{2}}.$ In the
view of above, we write invariant Yukawa Lagrangian for leptons and quarks.
The neutrinos mass terms can only arise from this invariant symmetric Yukawa
Lagrangian,%
\begin{equation}
-\pounds _{leptons}^{Y}=y_{\alpha \beta }^{e}\bar{l}_{_{\alpha }L}\eta
^{\ast }e_{\beta R}+y_{\alpha \beta }^{E}\bar{l}_{_{\alpha }L}\chi ^{\ast
}E_{\beta R}+y_{\alpha \beta }^{\nu }\bar{l}_{_{\alpha }L}\rho ^{\ast }\nu
_{\beta R}+\frac{1}{2}M_{R}\bar{\nu}_{\alpha R}^{C}\nu _{\alpha R}+h.c,
\label{Y3}
\end{equation}%
\begin{eqnarray}
-\pounds _{quarks}^{Y} &=&y_{i\alpha }^{d}\bar{Q}_{iL}\rho d_{\alpha
R}+y_{3\alpha }^{d}\bar{Q}_{3L}\eta ^{\ast }d_{\alpha R}+y_{i\alpha }^{u}%
\bar{Q}_{iL}\eta u_{\alpha R}+y_{3\alpha }^{u}\bar{Q}_{3L}\rho ^{\ast
}u_{\alpha R}  \notag \\
&&+y_{ij}^{J}\bar{Q}_{iL}\chi J_{jR}+y_{33}^{J}\bar{Q}_{3L}\chi ^{\ast
}J_{3R}+h.c.  \label{Y4}
\end{eqnarray}%
After SSB, the Yukawa interactions Eqs. (\ref{Y3} and \ref{Y4}) give the
following mass matrices for leptons and quarks in flavour basis.

\subsubsection{SM Charged Lepton and Neutrino Mass Matrices}

The Yukawa interaction term in Eq. (\ref{Y3}) reflects the charged leptons
mass matrix in flavour basis $(e,\mu ,\tau )$

\begin{equation}
M_{e}=-\frac{\upsilon _{\eta }}{\sqrt{2}}\left( 
\begin{array}{ccc}
y_{11}^{e} & y_{12}^{e} & y_{13}^{e} \\ 
y_{21}^{e} & y_{22}^{e} & y_{23}^{e} \\ 
y_{31}^{e} & y_{32}^{e} & y_{33}^{e}%
\end{array}%
\right) .  \label{Ye}
\end{equation}%
The masses of the SM charged leptons can be calculated using the appropriate
Yukawa Couplings and the diagonalization of the mass matrix $M_{e}$ as
follows;%
\begin{equation}
U_{L}^{e\ast }M_{e}U_{R}^{e\ast \dag }=diag(m_{e},m_{\mu },m_{\tau }),
\label{Ye1}
\end{equation}%
where with $U_{L,R}^{e\ast }~$ are unitary matrices that relate the flavour
eigenstate to the mass eigenstate (labelled with the superscript $\prime $)
of \ the SM as 
\begin{equation}
\left( 
\begin{array}{c}
e^{\prime } \\ 
\mu ^{\prime } \\ 
\tau ^{\prime }%
\end{array}%
\right) _{L,R}=U_{L,R}^{e\ast }\left( 
\begin{array}{c}
e \\ 
\mu \\ 
\tau%
\end{array}%
\right) _{L,R}.  \label{Y7a}
\end{equation}

The Dirac $y_{\alpha \beta }^{\nu }\bar{l}_{_{\alpha }L}\rho ^{\ast }\nu
_{\beta R}$ and Majorana $\frac{1}{2}M_{R}\bar{\nu}_{\alpha R}^{C}\nu
_{\alpha R}$ mass term in Yukawa interaction Eq. (\ref{Y3}) gives $6\times 6$
mass matrix of neutrinos in flavour basis $(\nu _{eL},\nu _{\mu L},\nu
_{\tau L},\nu _{eR}^{C},\nu _{\mu R}^{C},\nu _{\tau R}^{C})$;

\begin{equation}
M_{v}=\left( 
\begin{array}{cc}
0 & M_{D} \\ 
M_{D}^{T} & -M_{R}%
\end{array}%
\right) ,  \label{Y8}
\end{equation}%
where $M_{D}=\frac{\upsilon _{\rho }}{\sqrt{2}}y_{\alpha \beta }^{\nu }$ and 
$M_{R}=M~$are the mass parameters for Dirac and RH neutrinos respectively. $%
y_{\alpha \beta }^{\nu }~$is a $3\times 3$ complex Yukawa coupling matrix,
and $M$ is a $3\times 3$ diagonal and degenerate mass matrix for RH
neutrino. When taking into account the fact that all of the $M_{R}$'s
eigenvalues are greater than the $M_{D}$'s eigenvalues, we get the masses of
the left and RH neutrinos after diagonalizing $6\times 6$ neutrino mass
matrix $M_{\nu }$;%
\begin{equation}
U^{v\ast }M_{\nu }U^{v\ast \dag }=diag(m_{\nu L},m_{\nu R}).  \label{Y9}
\end{equation}%
The order of magnitude of neutrino's mass is%
\begin{equation}
m_{\nu L}\approx \frac{\upsilon _{\rho }^{2}}{M},~m_{\nu R}\approx M_{R}.
\label{Y10}
\end{equation}%
The flavour eigenstate related with mass eigenstate (labelled with the
superscript $\prime $) given as;%
\begin{equation}
\left( 
\begin{array}{c}
\nu _{L}^{\prime } \\ 
\nu _{R}^{\prime }%
\end{array}%
\right) =\left( 
\begin{array}{cc}
U_{L}^{v\ast } & U_{LR}^{v\ast } \\ 
U_{RL}^{v\ast } & U_{R}^{v\ast }%
\end{array}%
\right) \left( 
\begin{array}{c}
\nu _{L} \\ 
\nu _{R}%
\end{array}%
\right) ,  \label{Y11}
\end{equation}%
where $U_{L}^{\nu \ast }$ is standard LH neutrino mixing matrix, $U_{R}^{\nu
\ast }$ is RH heavy neutrino mixing matrix, the mixing and $U_{LR}^{\nu \ast
}$ represent the mixing among LH and RH neutrinos. Here the mixing matrix
for charged leptons $U_{LR}^{e\ast }~$in Eq. (\ref{Y7a}) and mixing matrix
of LH light neutrino $U_{L}^{\nu \ast }$allows us to define
Pontecorvo--Maki--Nakagawa--Sakata (PMNS) matrix's definition%
\begin{equation}
U_{PMNS}=U_{L}^{\nu \ast }U_{L}^{e\ast \dag }.  \label{PMNS}
\end{equation}%
Here the flavor and mass eigenstate of the SM charged leptons taken as the
same $(U_{L}^{e\ast }=I)$. Whereas, all the rotations introduced in
neutrinos are represented by $U_{PMNS}=U_{L}^{\nu \ast }.$

Above, the canonical type-I seesaw mechanism \cite{seesaw-I,seesaw-Ia} has
been used to generate small neutrino masses, we can see in Eq. (\ref{Y10}),
large values of RH neutrino masses produce light neutrino masses. Here we
recall the Eq. (\ref{Y1}) which shows that the theory possesses a Landau
pole. It also yields a constraint on the mixing angle $\sin ^{2}\theta
_{W}<0.25$ that keeps the theory within the perturbative regime and compels
us to select a maximum value of $M$ of about $5TeV$ \cite%
{331-model-energy-scale1,331-model-energy-scale2}. Consequently, we require $%
\upsilon _{\rho }\approx 10^{-2}GeV$ in order to obtain LH neutrino's masses
at the $eV$ scale. As we can see in Eq. (\ref{G11a}), there is no lower
limit on the VEV of $\upsilon _{\rho }$, only upper limit of $\approx 246GeV$
is valid. Therefore, we can reduce $\upsilon _{\rho }$ to the extent we like.

\subsubsection{SM Up and Down Type Quarks Mass Matrices}

One of the scalars in the scalar spectrum should acquire a small VEV $%
\upsilon _{\rho }$, caution is necessary to assess whether the calculated
quark masses can be derived from this spectrum. Fortunately, Yukawa
interactions involving both the scalar triplet $\rho $ and $\eta $ serve as
sources for both up and down-type SM quark masses. This is sufficient to
ensure that all SM quark masses are produced correctly when a small VEV $%
\upsilon _{\rho }$ and a regular $\upsilon _{\eta }$ are combined.

In Yukawa interaction Eq. (\ref{Y4}), the terms$\ y_{i\alpha }^{u}\bar{Q}%
_{iL}\eta u_{\alpha R}$ and $y_{3\alpha }^{u}\bar{Q}_{3L}\rho ^{\ast
}u_{\alpha R}$ give mass matrix for up-type quarks whereas $y_{i\alpha }^{d}%
\bar{Q}_{iL}\rho d_{\alpha R}$ and $y_{3\alpha }^{d}\bar{Q}_{3L}\eta ^{\ast
}d_{\alpha R}$ give mass matrix for down-type quarks in flavor basis $%
(u,c,t) $ and $(d,s,b)$ respectively; 
\begin{eqnarray}
M^{u} &=&\frac{1}{\sqrt{2}}\left( 
\begin{array}{ccc}
-\upsilon _{\eta }y_{11}^{u} & -\upsilon _{\eta }y_{12}^{u} & -\upsilon
_{\eta }y_{13}^{u} \\ 
-\upsilon _{\eta }y_{21}^{u} & -\upsilon _{\eta }y_{22}^{u} & -\upsilon
_{\eta }y_{23}^{u} \\ 
\upsilon _{\rho }y_{31}^{u} & \upsilon _{\rho }y_{32}^{u} & \upsilon _{\rho
}y_{33}^{u}%
\end{array}%
\right) ,  \label{Yu1} \\
M^{d} &=&-\frac{1}{\sqrt{2}}\left( 
\begin{array}{ccc}
\upsilon _{\rho }y_{11}^{d} & \upsilon _{\rho }y_{12}^{d} & \upsilon _{\rho
}y_{13}^{d} \\ 
\upsilon _{\rho }y_{21}^{d} & \upsilon _{\rho }y_{22}^{d} & \upsilon _{\rho
}y_{23}^{d} \\ 
\upsilon _{\eta }y_{31}^{d} & \upsilon _{\eta }y_{32}^{d} & \upsilon _{\eta
}y_{33}^{d}%
\end{array}%
\right) .  \label{Yd1}
\end{eqnarray}%
After suitable selection of Yukawa coupling and diagonalization of mass
matrices $M_{u}$ and $M_{d},$ we can obtain the masses of SM quarks as;%
\begin{eqnarray}
U_{L}^{u\ast }M_{u}U_{R}^{u\ast \dag } &=&diag(m_{u},m_{c},m_{t})\text{ and}
\label{Yu2} \\
U_{L}^{d\ast }M_{d}U_{R}^{d\ast \dag } &=&diag(m_{d},m_{s},m_{b}),
\label{Yd2}
\end{eqnarray}%
where with $U_{L,R}^{u\ast }~$and $V_{L,R}^{d\ast }$ are unitary matrices
that relate flavour eigenstate to the mass eigenstate (superscript $\prime $%
) of up-type and down-type quarks respectively given as%
\begin{eqnarray}
\left( 
\begin{array}{c}
u^{\prime } \\ 
c^{\prime } \\ 
t^{\prime }%
\end{array}%
\right) _{L,R} &=&U_{L,R}^{u\ast }\left( 
\begin{array}{c}
u \\ 
c \\ 
t%
\end{array}%
\right) _{L,R}\text{ and}  \label{Yu3} \\
\left( 
\begin{array}{c}
d^{\prime } \\ 
s^{\prime } \\ 
b^{\prime }%
\end{array}%
\right) _{L,R} &=&U_{L,R}^{d\ast }\left( 
\begin{array}{c}
d \\ 
s \\ 
b%
\end{array}%
\right) _{L,R},  \label{Yd3}
\end{eqnarray}%
it allows for the Cabibbo-Kobayashi-Maskawa (CKM) matrix's definition \cite%
{C5}:%
\begin{equation}
V_{CKM}=U_{L}^{u\ast }U_{L}^{d\ast \dag }.  \label{CKM}
\end{equation}%
Only the LH quarks rotation matrix is introduced in the SM. Due both
gauge-singlet characteristics and the universality of the fermion-gauge
coupling, the rotation matrices of the RH fermions can be absorbed by the
redefinition of fermion fields. It is worth mentioning that in the SM,
usually the flavour eigenstate and mass eigenstate of up-type quarks are
taken as the same $(U_{L}^{\ast }=I)$, and all the rotation is introduced in
down-type quarks $(V_{CKM}=U_{L}^{d\ast \dag })$ whereas in the 3-3-1 model,
rotation matrices $U_{L}^{\ast }$ and $U_{L}^{d\ast }$ each have a distinct
physical meaning. As the LH quarks nonuniversaly couples with the $Z_{0}$
boson, therefore we get the distinct representations of the three SM LH
quarks favours.

\subsubsection{New Heavy Exotic Leptons and Quarks Mass Matrices}

The term $y_{\alpha \beta }^{E}\bar{l}_{_{\alpha }L}\chi ^{\ast }E_{\beta R}$
and $y_{ij}^{J}\bar{Q}_{iL}\chi J_{jR}~$in Yukawa interaction Eqs. (\ref{Y3}
and \ref{Y4}) gives heavy exotic leptons and quarks mass matrices on flavour
basis $(E_{e},E_{\mu },E_{\tau })~$and $\left( J_{1},J_{2}\right) $
respectively;

\begin{equation}
M_{E}=-\frac{\upsilon _{\chi }}{\sqrt{2}}\left( 
\begin{array}{ccc}
y_{11}^{E} & y_{12}^{E} & y_{13}^{E} \\ 
y_{21}^{E} & y_{22}^{E} & y_{23}^{E} \\ 
y_{31}^{E} & y_{32}^{E} & y_{33}^{E}%
\end{array}%
\right) ,  \label{EE}
\end{equation}%
\begin{equation}
M_{J}=-\frac{\upsilon _{\chi }}{\sqrt{2}}\left( 
\begin{array}{cc}
y_{11}^{J} & y_{12}^{J} \\ 
y_{21}^{J} & y_{22}^{J}%
\end{array}%
\right) .  \label{JJ}
\end{equation}

With the suitable selection of Yukawa couplings and the diagonalization of
mass matrices $M_{E}$ and $M_{J}$, the masses of heavy exotic charged
leptons and quarks can be obtained as: 
\begin{eqnarray}
U_{L}^{E\ast }M_{E}U_{R}^{E\ast \dag } &=&diag(m_{E_{e}},m_{E_{\mu
}},m_{E_{\tau }}),  \label{mEE} \\
U_{L}^{J\ast }M_{J}U_{R}^{J\ast \dag } &=&diag(m_{J_{1}},m_{J_{2}}),
\label{mJ}
\end{eqnarray}%
where $U_{L,R}^{E\ast }$ and $U_{L,R}^{J\ast }~$are unitary matrices that
relate flavour eigenstate to the mass eigenstate (superscript $\prime $) of
heavy leptons and quarks given as 
\begin{eqnarray}
\left( 
\begin{array}{c}
E_{e}^{\prime } \\ 
E_{\mu }^{\prime } \\ 
E_{\tau }^{\prime }%
\end{array}%
\right) _{L,R} &=&U_{L,R}^{E\ast }\left( 
\begin{array}{c}
E_{e} \\ 
E_{\mu } \\ 
E_{\tau }%
\end{array}%
\right) _{L,R},  \label{UEE} \\
\left( 
\begin{array}{c}
J_{1}^{\prime } \\ 
J_{2}^{\prime }%
\end{array}%
\right) _{L,R} &=&U_{L,R}^{J\ast }\left( 
\begin{array}{c}
J_{1} \\ 
J_{2}%
\end{array}%
\right) _{L,R}.  \label{UJ}
\end{eqnarray}%
The exotic quarks $J_{3}$ cannot be mixed with $J_{1}$ and $J_{2}~$because
they have different charges$.$ For the sake of simplicity, we take flavour
eigenstate to be the same as mass eigenstate of new heavy exotic leptons and
quarks ($U_{L,R}^{E\ast }=I_{3\times 3}~$and $U_{L,R}^{J\ast }=I_{2\times 2}$%
).

\subsection{Higgs Potential}

The renormalizable gauge $SU(3)_{C}\otimes SU(3)_{L}\otimes U(1)_{X}$ and $%
Z_{4}$ symmetry invariant potential for the scalar content can be written as:%
\begin{eqnarray}
-V_{Higgs} &=&\mu _{1}^{2}\left( \rho \rho ^{\dag }\right) +\mu
_{2}^{2}\left( \eta \eta ^{\dag }\right) +\mu _{3}^{2}\left( \chi \chi
^{\dag }\right) +\lambda _{1}\left( \rho \rho ^{\dag }\right) ^{2}+\lambda
_{2}\left( \eta \eta ^{\dag }\right) ^{2}+\lambda _{3}\left( \chi \chi
^{\dag }\right) ^{2}  \notag \\
&&+\lambda _{12}\left( \rho \rho ^{\dag }\right) \left( \eta \eta ^{\dag
}\right) +\lambda _{13}\left( \rho \rho ^{\dag }\right) \left( \chi \chi
^{\dag }\right) ++\lambda _{23}\left( \eta \eta ^{\dag }\right) \left( \chi
\chi ^{\dag }\right)  \notag \\
&&+\lambda _{12}^{\prime }\left( \rho ^{\dag }\eta \right) \left( \eta
^{\dag }\rho \right) +\lambda _{13}^{\prime }\left( \rho ^{\dag }\chi
\right) \left( \chi ^{\dag }\rho \right) +\lambda _{23}^{\prime }\left( \eta
^{\dag }\chi \right) \left( \chi ^{\dag }\eta \right)  \notag \\
&&+\sqrt{2}f\left( \epsilon _{ijk}\rho ^{i}\eta ^{j}\chi ^{k}+h.c\right) ,
\label{H1}
\end{eqnarray}%
where $\lambda _{i(j)}^{(^{\prime })}$'s are dimensionless parameters. The
parameters $\mu _{i}$ and $f$ have the dimension of mass. For simplicity we
consider $f=k$ $\upsilon _{\chi }$ with $k\sim O(10^{-3)}$. The other terms
i.e., $(\chi ^{\dag }\eta \rho ^{\dag }\eta +h.c)$ \cite{scalar-sector} are
invariant under $SU(3)_{C}\otimes SU(3)_{L}\otimes U(1)_{X}$ gauge but
violate $Z_{4}$ symmetry, therefore these terms are not introduced in scalar
potential. The neutral component of scalar fields can be shifted as follow 
\cite{C1.2}:%
\begin{eqnarray}
\rho ^{0} &=&\frac{1}{\sqrt{2}}\left( \upsilon _{\rho }+\xi _{\rho }+i\zeta
_{\rho }\right) ,  \notag \\
\eta ^{0} &=&\frac{1}{\sqrt{2}}\left( \upsilon _{\eta }+\xi _{\eta }+i\zeta
_{\eta }\right) ,  \notag \\
\chi ^{0} &=&\frac{1}{\sqrt{2}}\left( \upsilon _{\chi }+\xi _{\chi }+i\zeta
_{\chi }\right) .  \label{H1c}
\end{eqnarray}

Taking the derivative of Higgs potential Eq. (\ref{H1}) with respect to
scalar field at global minima \cite{Higgs-mass-matrix}$,$ 
\begin{equation}
\left \vert \frac{\partial V_{Higgs}}{\partial \rho }\right \vert 
_{\substack{ \rho =\left \langle \rho \right \rangle =\upsilon _{\rho }  \\ %
\eta =\left \langle \eta \right \rangle =\upsilon _{\eta }  \\ \chi =\left
\langle \chi \right \rangle =\upsilon _{\chi }}}=\left \vert \frac{\partial
V_{Higgs}}{\partial \eta }\right \vert _{\substack{ \rho =\left \langle \rho
\right \rangle =\upsilon _{\rho }  \\ \eta =\left \langle \eta \right
\rangle =\upsilon _{\eta }  \\ \chi =\left \langle \chi \right \rangle
=\upsilon _{\chi }}}=\left \vert \frac{\partial V_{Higgs}}{\partial \chi }%
\right \vert _{\substack{ \rho =\left \langle \rho \right \rangle =\upsilon
_{\rho }  \\ \eta =\left \langle \eta \right \rangle =\upsilon _{\eta }  \\ %
\chi =\left \langle \chi \right \rangle =\upsilon _{\chi }}}=0,  \label{H2}
\end{equation}%
gives the following tadpole constraint equations: 
\begin{eqnarray}
\mu _{1}^{2}\upsilon _{\rho }+\lambda _{1}\upsilon _{\rho }^{3}+\frac{1}{2}%
\lambda _{12}\upsilon _{\rho }\upsilon _{\eta }^{2}+\frac{1}{2}\lambda
_{13}\upsilon _{\rho }\upsilon _{\chi }^{2} &=&f\upsilon _{\eta }\upsilon
_{\chi },  \label{H3a} \\
\mu _{2}^{2}\upsilon _{\eta }+\lambda _{2}\upsilon _{\eta }^{3}+\frac{1}{2}%
\lambda _{12}\upsilon _{\eta }\upsilon _{\rho }^{2}+\frac{1}{2}\lambda
_{23}\upsilon _{\eta }\upsilon _{\chi }^{2} &=&f\upsilon _{\rho }\upsilon
_{\chi },  \label{H3b} \\
\mu _{3}^{2}\upsilon _{\chi }+\lambda _{3}\upsilon _{\chi }^{3}+\frac{1}{2}%
\lambda _{13}\upsilon _{\chi }\upsilon _{\rho }^{2}+\frac{1}{2}\lambda
_{23}\upsilon _{\chi }\upsilon _{\eta }^{2} &=&f\upsilon _{\rho }\upsilon
_{\eta }.  \label{H3}
\end{eqnarray}%
The first Eq. (\ref{H3a}) of the three aforementioned equations is the one
that matters to us because it will provide the desired relation for a small
value for $\upsilon _{\rho }$, which is necessary for the production of tiny
neutrino masses. If we want to remain in the perturbative domain, as we have
already explained in the begning of section 3.3, the maximum energy scale in
this version of 3-3-1 model $M$ is approximately $5TeV$. We relate the
scalar $\rho $ to this scale, so that $\mu _{1}\approx $ $M$. If we take
note of the fact that $\mu _{1}$ becomes dominant on the left in the Eq. (%
\ref{H3a}). we get: 
\begin{equation}
\upsilon _{\rho }\simeq \frac{f\upsilon _{\eta }\upsilon _{\chi }}{\mu
_{1}^{2}}=\frac{f\upsilon _{\eta }\upsilon _{\chi }}{M^{2}},  \label{H4}
\end{equation}%
this is the type II seesaw mechanism's feature \cite%
{Type-II-seesaw-331,Type-II-seesaw-331a}. we can also define $\tan \beta $
from Eq. (\ref{H4}),%
\begin{equation}
\frac{\upsilon _{\rho }}{\upsilon _{\eta }}=\frac{f\upsilon _{\chi }}{M^{2}}%
=\tan \beta .  \label{H5}
\end{equation}

By putting these tadpole constraint Eqs. (\ref{H3a}, \ref{H3b} and \ref{H3})
into Higgs potential, we can get the masses and physical states of all
neutral and charged Higgs bosons.

\subsubsection{Neutral Higgs}

A scalar Higgs mass matrix can be constructed from Higgs potential, when the
real components $\xi _{\rho }$, $\xi _{\eta }$, and $\xi _{\chi }$ of
neutral scalars develop the VEVs By taking second derivative of Higgs
potential such that $M_{\Phi _{i}\Phi _{j}}^{2}=\frac{\partial ^{2}V_{Higgs}%
}{\partial \Phi _{i}\partial \Phi _{j}}\rfloor _{\Phi _{i}=0}$ with $\Phi
_{i}=$ $\xi _{\rho }$, $\xi _{\eta }$, $\xi _{\chi }$ \cite{scalar-sector}.
The square mass matrix of neutral scalar Higgs boson can be obtained and
written as, 
\begin{equation}
M_{H}^{2}=\left( 
\begin{array}{ccc}
2\lambda _{1}\upsilon _{\rho }^{2}+\frac{f\upsilon _{\eta }\upsilon _{\chi }%
}{\upsilon _{\rho }} & \lambda _{12}\upsilon _{\rho }\upsilon _{\eta
}-f\upsilon _{\chi } & \lambda _{13}\upsilon _{\rho }\upsilon _{\chi
}-f\upsilon _{\eta } \\ 
\lambda _{12}\upsilon _{\rho }\upsilon _{\eta }-f\upsilon _{\chi } & 
2\lambda _{2}\upsilon _{\eta }^{2}+\frac{f\upsilon _{\rho }\upsilon _{\chi }%
}{\upsilon _{\eta }} & \lambda _{23}\upsilon _{\eta }\upsilon _{\chi
}-f\upsilon _{\rho } \\ 
\lambda _{13}\upsilon _{\rho }\upsilon _{\chi }-f\upsilon _{\eta } & \lambda
_{23}\upsilon _{\eta }\upsilon _{\chi }-f\upsilon _{\rho } & 2\lambda
_{3}\upsilon _{\chi }^{2}+\frac{f\upsilon _{\rho }\upsilon _{\eta }}{%
\upsilon _{\chi }}%
\end{array}%
\right) .  \label{H8}
\end{equation}

In the limit $\upsilon _{\chi }>>\upsilon _{\rho },\upsilon _{\eta }$, there
are three eigenvalues of scalar Higgs mass matrix Eq. (\ref{H8}). One
eigenvalue represents SM Higgs ($h$) and other two are heavy Higgs bosons
named $H_{2}$ and $H_{3}$\cite{Collider-Phenomenology}. 
\begin{equation}
m_{h}^{2}=\frac{2(\lambda _{1}\upsilon _{\rho }^{4}+\lambda _{12}\upsilon
_{\rho }^{2}\upsilon _{\eta }^{2}+\lambda _{2}\upsilon _{\eta }^{4})}{%
\upsilon _{\eta }^{2}+\upsilon _{\rho }^{2}},\text{~}M_{H_{2}}^{2}=\frac{%
\upsilon _{\eta }^{2}+\upsilon _{\rho }^{2}}{\upsilon _{\rho }\upsilon
_{\eta }}f\upsilon _{\chi },~M_{H_{3}}^{2}=2\lambda _{3}\upsilon _{\chi
}^{2}.  \label{H9}
\end{equation}

The mass eigenstates $(h,~H_{2},~H_{3})$ of scalar Higgs are related \ to
weak eigenstates $(\xi _{\rho },~\xi _{\eta },~\xi _{\chi })$ by a rotation
given as \cite{Collider-Phenomenology}, 
\begin{equation}
\left( 
\begin{array}{c}
h \\ 
H_{2} \\ 
H_{3}%
\end{array}%
\right) =\left( 
\begin{array}{ccc}
\cos \beta _{12} & -\sin \beta _{12} & 0 \\ 
\sin \beta _{12} & \cos \beta _{12} & 0 \\ 
0 & 0 & 1%
\end{array}%
\right) ^{\dag }\left( 
\begin{array}{c}
\xi _{\rho } \\ 
\xi _{\eta } \\ 
\xi _{\chi }%
\end{array}%
\right) ,  \label{H10}
\end{equation}%
where $\sin \beta _{12}=\frac{\upsilon _{\eta }}{\sqrt{\upsilon _{\eta
}^{2}+\upsilon _{\rho }^{2}}}$ and $\cos \beta _{12}=\frac{\upsilon _{\rho }%
}{\sqrt{\upsilon _{\eta }^{2}+\upsilon _{\rho }^{2}}}.$

When the imaginary components ($\zeta _{\rho },~\zeta _{\eta },~\zeta _{\chi
}$) of neutral scalar develop the VEVs, the Higgs potential gives rise to
the mass matrix of pseudoscalar Higgs bosons. By taking second derivative of
Higgs potential such that $M_{\Phi _{i}\Phi _{j}}^{2}=\frac{\partial
^{2}V_{Higgs}}{\partial \Phi _{i}\partial \Phi _{j}}\rfloor _{\Phi _{i}=0}$
with $\Phi _{i}=$ $\zeta _{\rho },~\zeta _{\eta },~\zeta _{\chi }$ \cite%
{scalar-sector} we can get square mass matrix of pseudoscalar Higgs and
written as,%
\begin{equation}
M_{H^{\prime }}^{2}=\left( 
\begin{array}{ccc}
\frac{\upsilon _{\eta }\upsilon _{\chi }}{\upsilon _{\rho }} & \upsilon
_{\chi } & \upsilon _{\eta } \\ 
\upsilon _{\chi } & \frac{\upsilon _{\rho }\upsilon _{\chi }}{\upsilon
_{\eta }} & \upsilon _{\rho } \\ 
\upsilon _{\eta } & \upsilon _{\rho } & \frac{\upsilon _{\rho }\upsilon
_{\eta }}{\upsilon _{\chi }}%
\end{array}%
\right) f.  \label{H11}
\end{equation}

There are three eigenvalues of the pseudoscalar Higgs mass matrix Eq. (\ref%
{H11}). One non-zero eigenvalue represents the pseudoscalar Higgs $H_{0}$
while other two zero eigenvalues represent the goldstone bosons $G_{Z}~$and $%
G_{Z^{\prime }}$ eaten up by $Z$ and $Z^{\prime }$ gauge bosons \cite%
{Collider-Phenomenology}. 
\begin{equation}
M_{H_{0}}^{2}=\frac{(\upsilon _{\rho }^{2}\upsilon _{\eta }^{2}+\upsilon
_{\eta }^{2}\upsilon _{\chi }^{2}+\upsilon _{\rho }^{2}\upsilon _{\chi
}^{2})f}{\upsilon _{\rho }\upsilon _{\eta }\upsilon _{\chi }},\text{~}%
M_{G_{Z}}^{2}=0,~M_{G_{Z^{\prime }}}^{2}=0.  \label{H12}
\end{equation}

The mass eigenstates $(H_{0},~G_{Z},~G_{Z^{\prime }})~$of pseudoscalar Higgs
are related to with weak eigenstates $(\zeta _{\rho },~\zeta _{\eta },~\zeta
_{\chi })$ by a rotation given as \cite{Collider-Phenomenology}, 
\begin{equation}
\left( 
\begin{array}{c}
H_{0} \\ 
G_{Z} \\ 
G_{Z^{\prime }}%
\end{array}%
\right) =\left( 
\begin{array}{ccc}
\sin \beta _{12} & \cos \beta _{12} & 0 \\ 
-\cos \beta _{12} & \sin \beta _{12} & 0 \\ 
0 & 0 & 1%
\end{array}%
\right) ^{\dag }\left( 
\begin{array}{c}
\zeta _{\rho } \\ 
\zeta _{\eta } \\ 
\zeta _{\chi }%
\end{array}%
\right) .  \label{H13}
\end{equation}

\subsubsection{Charged Higgs}

In addition to the neutral Higgs bosons as previously discussed, there are
two types of charged Higgs bosons with electric charges of $\pm 1$and $\pm 2$%
. The square mass matrix of charged Higgs bosons can be obtained by taking
second derivative of Higgs potential such that $M_{\Phi _{i}^{\ast }\Phi
_{j}}^{2}=\frac{\partial ^{2}V_{Higgs}}{\partial \Phi _{i}^{\ast }\partial
\Phi _{j}}\rfloor _{\Phi _{i}=0}$ with $\Phi _{i}=$ $\eta ^{\pm },~\rho
^{\pm },$ $\chi ^{\mp }{}^{^{\ast }},\eta ^{\mp }{}^{^{\ast }}$(for charge $%
\pm 1)$ and $\Phi _{i}=$ $\chi ^{\pm 2},~\rho ^{\pm 2}{}($for charge $\pm 2)$
\cite{scalar-sector}. The square mass matrix of singly charged Higgs bosons
is%
\begin{equation}
M_{H^{\pm }}^{2}=\left( 
\begin{array}{cccc}
\frac{1}{2}\lambda _{12}^{\prime }\upsilon _{\eta }^{2}+\frac{f\upsilon
_{\eta }\upsilon _{\chi }}{\upsilon _{\rho }} & -\frac{1}{2}\lambda
_{12}^{\prime }\upsilon _{\rho }\upsilon _{\eta }-f\upsilon _{\chi } & 0 & 0
\\ 
-\frac{1}{2}\lambda _{12}^{\prime }\upsilon _{\rho }\upsilon _{\eta
}-f\upsilon _{\chi } & \frac{1}{2}\lambda _{12}^{\prime }\upsilon _{\rho
}^{2}+\frac{f\upsilon _{\rho }\upsilon _{\chi }}{\upsilon _{\eta }} & 0 & 0
\\ 
0 & 0 & \frac{1}{2}\lambda _{23}^{\prime }\upsilon _{\chi }^{2}+\frac{%
f\upsilon _{\rho }\upsilon _{\chi }}{\upsilon _{\eta }} & -\frac{1}{2}%
\lambda _{23}^{\prime }\upsilon _{\eta }\upsilon _{\chi }-f\upsilon _{\rho }
\\ 
0 & 0 & -\frac{1}{2}\lambda _{23}^{\prime }\upsilon _{\eta }\upsilon _{\chi
}-f\upsilon _{\rho } & \frac{1}{2}\lambda _{23}^{\prime }\upsilon _{\eta
}^{2}+\frac{f\upsilon _{\rho }\upsilon _{\eta }}{\upsilon _{\chi }}%
\end{array}%
\right) .  \label{H14}
\end{equation}

The diagonalization of $M_{H^{\pm }}^{2}$gives the following mass spectrum 
\cite{Collider-Phenomenology}:%
\begin{eqnarray}
M_{G_{W}^{\pm }} &=&0;~~M_{H_{1}^{\pm }}=\frac{\lambda _{12}^{\prime
}\upsilon _{\rho }\upsilon _{\eta }+2f\upsilon _{\chi }}{2\upsilon _{\rho
}\upsilon _{\eta }}\left( \upsilon _{\eta }^{2}+\upsilon _{\rho }^{2}\right)
,  \notag \\
M_{G_{W^{\prime }}^{\pm }} &=&0;~~M_{H_{2}^{\pm }}=\frac{\lambda
_{23}^{\prime }\upsilon _{\eta }\upsilon _{\chi }+2f\upsilon _{\rho }}{%
2\upsilon _{\eta }\upsilon _{\chi }}\left( \upsilon _{\eta }^{2}+\upsilon
_{\chi }^{2}\right) ,  \label{H15}
\end{eqnarray}%
where $H_{1}^{\pm }$ and $H_{2}^{\pm }$ are physical singly heavy charged
Higgs, where as $G_{W}^{\pm }$ and $G_{W^{\prime }}^{\pm }$ are two massless
goldstone bososn eaten by $W^{\pm }$ and $W^{^{\prime }\pm }$ gauge boson.
The mass eigenstates $(H_{1}^{\pm }~G_{W}^{\pm },~H_{2}^{\pm },G_{W^{\prime
}}^{\pm })~$of singly charged Higgs are related to weak eigenstates $(\eta
^{\pm },~\rho ^{\pm },~\chi ^{\mp ^{\ast }},\eta ^{\mp ^{\ast }})$ by a
rotation given as \cite{Collider-Phenomenology}

\begin{equation}
\left( 
\begin{array}{c}
H_{1}^{\pm } \\ 
G_{W}^{\pm } \\ 
H_{2}^{\pm } \\ 
G_{W^{\prime }}^{\pm }%
\end{array}%
\right) =\left( 
\begin{array}{cccc}
\cos \beta _{_{12}} & -\sin \beta _{_{12}} & 0 & 0 \\ 
\sin \beta _{_{12}} & \cos \beta _{_{12}} & 0 & 0 \\ 
0 & 0 & \cos \beta _{_{23}} & -\sin \beta _{23} \\ 
0 & 0 & \sin \beta _{23} & \cos \beta _{_{23}}%
\end{array}%
\right) \left( 
\begin{array}{c}
\eta ^{\pm } \\ 
\rho ^{\pm } \\ 
\chi ^{\mp ^{\ast }} \\ 
\eta ^{\mp }{}^{^{\ast }}%
\end{array}%
\right) ,  \label{H16}
\end{equation}%
where $\sin \beta _{23}=\frac{\upsilon _{\chi }}{\sqrt{\upsilon _{\eta
}^{2}+\upsilon _{\chi }^{2}}}$, $\cos \beta _{23}=\frac{\upsilon _{\eta }}{%
\sqrt{\upsilon _{\eta }^{2}+\upsilon _{\chi }^{2}}}~$and $\sin \beta
_{_{12}} $, $\cos \beta _{_{12}}$ already defined in neutral Higgs mixing
matrix.

The square mass matrix of doubly charged Higgs bosons is%
\begin{equation}
M_{H^{\pm 2}}^{2}=\left( 
\begin{array}{cc}
\frac{1}{2}\lambda _{13}^{\prime }\upsilon _{\chi }^{2}+\frac{f\upsilon
_{\eta }\upsilon _{\chi }}{\upsilon _{\rho }} & f\upsilon _{\eta }-\frac{1}{2%
}\lambda _{13}^{\prime }\upsilon _{\rho }\upsilon _{\chi } \\ 
f\upsilon _{\eta }-\frac{1}{2}\lambda _{13}^{\prime }\upsilon _{\rho
}\upsilon _{\chi } & \frac{1}{2}\lambda _{13}^{\prime }\upsilon _{\rho }^{2}+%
\frac{f\upsilon _{\rho }\upsilon _{\eta }}{\upsilon _{\eta }}%
\end{array}%
\right) .  \label{H17}
\end{equation}

The diagonalization of $M_{H^{\pm 2}}^{2}$ gives the following mass spectrum 
\cite{Collider-Phenomenology}%
\begin{equation}
M_{G_{V}^{\pm 2}}=0;~~M_{H^{\pm 2}}=\frac{\lambda _{13}^{\prime }\upsilon
_{\rho }\upsilon _{\chi }+2f\upsilon _{\eta }}{2\upsilon _{\rho }\upsilon
_{\chi }}\left( \upsilon _{\rho }^{2}+\upsilon _{\chi }^{2}\right) ,
\label{H18}
\end{equation}%
where $H^{\pm 2}$ is physical heavy doubly charged Higgs boson and $%
G_{V}^{\pm 2}$ is massless goldstone boson eaten up by $V^{\pm 2}$ heavy
doubly charged gauge boson. The mass eigenstates $(H^{\pm 2},~G_{V}^{\pm
2})~ $of doubly charged Higgs are related to weak eigenstates $(\chi ^{\pm
2},~\rho ^{\pm 2})$ by a rotation given as \cite{Collider-Phenomenology}

\begin{equation}
\left( 
\begin{array}{c}
H^{\pm 2} \\ 
G_{V}^{\pm 2}%
\end{array}%
\right) =\left( 
\begin{array}{cc}
\cos \beta _{13} & -\sin \beta _{_{13}} \\ 
\sin \beta _{_{13}} & \cos \beta _{13}%
\end{array}%
\right) \left( 
\begin{array}{c}
\chi ^{\pm 2} \\ 
\rho ^{\pm 2}%
\end{array}%
\right) ,  \label{H19}
\end{equation}%
where $\func{Si}n\beta _{13}=\frac{\upsilon _{\chi }}{\sqrt{\upsilon _{\rho
}^{2}+\upsilon _{\chi }^{2}}}$ and $Cos\beta _{13}=\frac{\upsilon _{\rho }}{%
\sqrt{\upsilon _{\rho }^{2}+\upsilon _{\chi }^{2}}}.$

\section{Input Parameters}

In Section 3.3.1, we previously talked about choosing a maximum value of $M$
of approximately $5TeV$ to maintain the theory in the perturbative regime.
By choosing the value of the first stage symmetry breaking scale $\upsilon
_{\chi }$, one can set the value of $f$ with the relation $f=k$ $\upsilon
_{\chi }$. The value of $\tan \beta $, $\upsilon _{\rho }$ and $\upsilon
_{\eta }$ can be fixed by using Eqs. (\ref{G11a} and \ref{H5}). The
parameters involved in the scalar sector are free parameters and can be
guessed with the eigenvalues of Higgs mass matrices. We summarize the values
of these parameters in Table \ref{parameters}.

%TCIMACRO{\TeXButton{B}{\begin{table}[tbp] \centering}}%
%BeginExpansion
\begin{table}[tbp] \centering%
%EndExpansion
\begin{tabular}{|c|c|}
\hline
\textbf{Model Parameter} & \textbf{Numerical Value} \\ \hline
$M$ & $5000~GeV$ \\ \hline
$\upsilon _{\chi }$ & $1000~GeV$ \\ \hline
$f$ & $3.0~GeV$ \\ \hline
$\tan \beta $ & $0.00012$ \\ \hline
$\upsilon _{\rho }$ & $0.0295~GeV$ \\ \hline
$\upsilon _{\eta }$ & $246~GeV$ \\ \hline
$\lambda _{1},\lambda _{2},\lambda _{3}$ & $0.134$ \\ \hline
$\lambda _{12},\lambda _{13}$ $and$ $\lambda _{23}$ & $0.5~and~0.05$ \\ 
\hline
$\lambda _{12}^{\prime },\lambda _{13}^{\prime },\lambda _{23}^{\prime }$ & $%
0.5$ \\ \hline
\end{tabular}%
%TCIMACRO{%
%\TeXButton{Parameters}{\caption{A possible set of input values of the 331 models' parameters \label{parameters}}}}%
%BeginExpansion
\caption{A possible set of input values of the 331 models' parameters \label{parameters}}%
%EndExpansion
%TCIMACRO{\TeXButton{E}{\end{table}}}%
%BeginExpansion
\end{table}%
%EndExpansion

\emph{By selecting a fair texture for the Yukawa matrices, one can determine
the right lepton and quark masses because Yukawa Couplings are free
parameters of the model. }

\begin{itemize}
\item \textbf{SM charged leptons Yukawa Coupling:} 
\begin{equation}
y_{\alpha \beta }^{e}=diag(-2.9350\times 10^{-6},~-6.0309\times
10^{-4},~-1.0201\times 10^{-2}).  \label{M1}
\end{equation}

\item \textbf{Neutrino Yukawa Coupling:}%
\begin{eqnarray}
y_{11}^{\nu } &=&2.9235\times 10^{-4},~y_{12}^{\nu }=5.4723\times
10^{-3},~y_{12}^{\nu }=3.4852\times 10^{-3},  \notag \\
y_{21}^{\nu } &=&-1.5811\times 10^{-4},~y_{22}^{\nu }=4.6532\times
10^{-3},~y_{23}^{\nu }=1.8294\times 10^{-2},  \notag \\
y_{31}^{\nu } &=&1.2465\times 10^{-4},~y_{32}^{\nu }=-6.9325\times
10^{-3},~y_{33}^{\nu }=1.5031\times 10^{-2}.  \label{M2}
\end{eqnarray}

\item \textbf{SM Up-Type Quarks Yukawa Coupling:}%
\begin{eqnarray}
y_{11}^{u} &=&4.06141\times 10^{-6},~y_{12}^{u}=-2.87185\times
10^{-6},~y_{12}^{u}=2.87185\times 10^{-6},  \notag \\
y_{21}^{u} &=&-4.93959\times 10^{-1},~y_{22}^{u}=-3.52728\times
10^{-1},~y_{23}^{u}=3.45835\times 10^{-1},  \notag \\
y_{31}^{u} &=&4.11632\times 10^{3},~y_{32}^{u}=2.88196\times
10^{3},~y_{33}^{u}=-2.9394\times 10^{3}.  \label{M3}
\end{eqnarray}

\item \textbf{SM Down-Type Quarks Yukawa Coupling:}%
\begin{eqnarray}
y_{11}^{d} &=&-6.28947\times 10^{-1},~y_{12}^{d}=-4.31217\times
10^{-1},~y_{12}^{d}=1.02082,  \notag \\
y_{21}^{d} &=&1.03437\times 10^{2},~y_{22}^{d}=1.03404\times
10^{2},~y_{23}^{d}=-3.29824,  \notag \\
y_{31}^{d} &=&1.11443\times 10^{-2},~y_{32}^{d}=1.14467\times
10^{-2},~y_{33}^{d}=3.65848\times 10^{-4}.  \label{M4}
\end{eqnarray}

\item \textbf{New Heavy Exotic Leptons and Quarks Yukawa Coupling:}%
\begin{equation*}
y_{\alpha \beta }^{E}=y_{\alpha \beta
}^{J}=diag(-1.13137,~-1.13137,~-1.13137).
\end{equation*}
\end{itemize}

\section{Numerical Results of Model from SPheno}

All the numerical values of model's parameters mentioned in the previous
topic have been used as input to the SPheno code generated from SARAH \cite%
{SARAH1}. And the masses of particles, mixing matrices, and flavour
violating observables have been obtained as output from the Spheno. The
numerical results of the model from Spheno and their observed values are
given below;

\begin{itemize}
\item \textbf{SM charged lepton and quarks masses;\ }%
\begin{eqnarray}
m_{e} &=&0.51099893~MeV,~m_{\mu }=105.658772~MeV,~m_{\tau }=1776.69~MeV, 
\notag \\
m_{u} &=&2.5~MeV,~m_{c}=1.27~GeV,~m_{t}=173.5~GeV,  \notag \\
m_{d} &=&5.0~MeV,~m_{s}=95.0~MeV,~m_{b}=4.18~GeV.  \label{N1}
\end{eqnarray}

\item \textbf{The up, down type quarks mixing matrices are given as;}
\end{itemize}

\begin{equation}
U_{L}^{u}=\left( 
\begin{array}{ccc}
-1.00000 & -3.16893\times 10^{-8} & 3.16886\times 10^{-8} \\ 
-4.48150\times 10^{-8} & 7.07107\times 10^{-1} & -7.07107\times 10^{-1} \\ 
5.28938\times 10^{-13} & -7.07107\times 10^{-1} & -7.07107\times 10^{-1}%
\end{array}%
\right) ,  \label{N1A}
\end{equation}%
\begin{equation}
U_{L}^{d}=\left( 
\begin{array}{ccc}
9.74951\times 10^{-1} & 1.53465\times 10^{-1} & -1.60993\times 10^{-1} \\ 
-2.22387\times 10^{-1} & 6.60308\times 10^{-1} & -7.17313\times 10^{-1} \\ 
-3.77719\times 10^{-3} & 7.35148\times 10^{-1} & 6.77896\times 10^{-1}%
\end{array}%
\right) .  \label{N1B}
\end{equation}

Using Eq. (\ref{CKM}) we get CKM matrix as; 
\begin{equation}
V_{CKM}=\left( 
\begin{array}{ccc}
-0.974951 & 0.222387 & 0.00377719 \\ 
0.222355 & 0.974125 & 0.0404829 \\ 
0.0053234 & 0.0403087 & -0.999173%
\end{array}%
\right) .  \label{N2}
\end{equation}

\begin{itemize}
\item \textbf{Observed values of charged leptons, quarks masses and CKM
matrix}$~$\cite{N}\textbf{:} 
\begin{eqnarray*}
m_{e} &=&0.5109989461\pm 0.0000000031~MeV, \\
m_{\mu } &=&105.6587745\pm 0.0000024~MeV, \\
m_{\tau } &=&1776.86\pm 0.12~MeV, \\
m_{u} &=&2.16_{-0.26}^{+0.49}~MeV,~1.27\pm 0.02~GeV,~m_{t}=172.76\pm
0.03~GeV, \\
m_{d}
&=&4.67_{-0.17}^{+0.48}~MeV,~m_{s}=93_{-5}^{+11}~MeV,~m_{b}=4.18_{-0.02}^{+0.03}~GeV,
\end{eqnarray*}
\end{itemize}

\begin{equation*}
V_{CKM}^{Observed}=\left( 
\begin{array}{ccc}
0.97401\pm 0.00011 & 0.22650\pm 0.00048 & 0.00361_{-0.00009}^{+0.00011} \\ 
0.22636\pm 0.00048 & 0.97320\pm 0.00011 & 0.04053_{-0.00061}^{+0.00083} \\ 
0.00854_{-0.00016}^{+0.00023} & 0.03978_{-0.00060}^{+0.00082} & 
0.999172_{-0.000035}^{+0.000024}%
\end{array}%
\right) .
\end{equation*}

\begin{itemize}
\item \textbf{Neutrino masses and PMNS matrix:}
\end{itemize}

The SPheno code has generated six mass eigenvalues and $6\times 6$ neutrino
mixing matrix $U^{\nu }$ after diagonalization neutrino mass matrix $M_{\nu
} $ Eq. (\ref{Y8}). Three of the eigenvalues correspond to the masses of
light LH neutrinos, whereas the other three correspond to the masses of
heavy RH neutrinos. The standard LH neutrinos masses have been obtained: 
\begin{equation}
m_{1}=2.32165\times 10^{-5}~eV,~m_{2}=8.43831\times
10^{-3}~eV,~m_{3}=5.11591\times 10^{-2}~eV.  \label{N2A}
\end{equation}%
We can obtain the following mass squared differences of light LH neutrinos
from Eq. (\ref{N2A}) 
\begin{eqnarray}
\Delta m_{21}^{2} &=&m_{2}^{2}-m_{1}^{2}=7.12\times 10^{-5}~eV^{2},  \notag
\\
\Delta m_{31}^{2} &=&m_{3}^{2}-m_{2}^{2}=2.55\times 10^{-3}~eV^{2}.
\label{N4}
\end{eqnarray}

The result of $6\times 6$ neutrino mixing matrix $U^{\nu }$ is%
\begin{equation*}
U^{\nu }=\left( 
\begin{array}{cccccc}
-0.816 & 0.559 & 0.146 & -2.99\times 10^{-9} & 2.27\times 10^{-8} & 
1.46\times 10^{-8} \\ 
0.453 & 0.461 & 0.763 & -4.22\times 10^{-9} & 1.89\times 10^{-8} & 
7.64\times 10^{-8} \\ 
-0.359 & -0.689 & 0.629 & 5.83\times 10^{-9} & -2.83\times 10^{-8} & 
6.28\times 10^{-8} \\ 
1.48\times 10^{-9} & -1.99\times 10^{-11} & -2.49\times 10^{-12} & 0.983 & 
0.184 & 0.0 \\ 
-5.60\times 10^{-10} & -4.17\times 10^{-8} & 4.88\times 10^{-11} & -0.184 & 
0.983 & 0.0 \\ 
-1.69\times 10^{-10} & -1.15\times 10^{-10} & -1.0\times 10^{-7} & 
-5.7\times 10^{-22} & 2.3\times 10^{-21} & 1.0%
\end{array}%
\right) .
\end{equation*}

The values of the standard LH neutrino mixing matrix $U_{L}^{\nu }$ have
been represented in the above $U^{\nu }$ matrix's left upper $3\times 3$
block as:%
\begin{equation}
U_{L}^{\nu }=U_{PMNS}=\left( 
\begin{array}{ccc}
-0.816 & 0.559 & 0.146 \\ 
0.453 & 0.461 & 0.763 \\ 
-0.359 & -0.689 & 0.629%
\end{array}%
\right) .  \label{N5}
\end{equation}

The mixing angles shown below have been calculated from the mixing matrix $%
U_{L}^{\nu }$%
\begin{equation}
\sin ^{2}\theta _{12}=0.304,~\sin ^{2}\theta _{23}=0.595,~\sin ^{2}\theta
_{13}=2.15\times 10^{-2}.  \label{N6}
\end{equation}

\begin{itemize}
\item \textbf{Observed values of the light neutrino's parameters}
\end{itemize}

The observed mass squared differences and mixing angles \cite{N}%
\begin{eqnarray*}
\Delta m_{21}^{2} &=&\left( 7.53\pm 0.18\right) \times 10^{-5}eV^{2}, \\
\Delta m_{32}^{2} &=&\left( -2.536\pm 0.034\right) \times 10^{-3}eV^{2}~\
(Inverted~Order), \\
\Delta m_{32}^{2} &=&\left( 2.453\pm 0.033\right) \times 10^{-3}eV^{2}~\  \  \
(Normal~Order), \\
\sin ^{2}\theta _{12} &=&0.307\pm 0.013, \\
\sin ^{2}\theta _{23} &=&0.539\pm 0.022~\  \  \ (Inverted~Order), \\
\sin ^{2}\theta _{23} &=&0.546\pm 0.021~\  \  \ (Normal~Order), \\
\sin ^{2}\theta _{13} &=&\left( 2.20\pm 0.07~\right) \times 10^{-2}.
\end{eqnarray*}

The Best fit value of PMNS matrix with $3\sigma ~(99.7~\%CL)$ range is \cite%
{PMNS}%
\begin{equation*}
U_{PMNS}^{Best~fit}=\left( 
\begin{array}{ccc}
0.801...0.845 & 0.513...0.579 & 0.143...0.156 \\ 
0.232...0.507 & 0.459...0.694 & 0.629...0.779 \\ 
0.260...0.526 & 0.470...0.702 & 0.702...0.763%
\end{array}%
\right) .
\end{equation*}

\begin{itemize}
\item \textbf{Gauge and Higgs Bosons Masses}
\end{itemize}

Table \ref{GH} provides the mass spectrum of the SM gauge boson and the
Higgs boson with observed values, whereas Table \ref{HGH} newly heavy gauge
and Higgs bosons have been calculated by our SPheno algorithm.

%TCIMACRO{\TeXButton{B}{\begin{table}[tbp] \centering}}%
%BeginExpansion
\begin{table}[tbp] \centering%
%EndExpansion
\begin{tabular}{|c|c|c|}
\hline
\multicolumn{3}{|c|}{\textbf{Mass Spectrum of SM Gauge and Higgs Boson}} \\ 
\hline
Particle & Predicted Value & Observed Value \\ \hline
$m_{Z}$ & $91.189~GeV$ & $91.187\pm 0.0021~GeV~$\cite{N} \\ \hline
$m_{W^{\pm }}$ & $80.350~GeV$ & $80.379\pm 0.012~GeV~$\cite{N} \\ \hline
$m_{h}$ & $125.09~GeV$ & $125.10\pm 0.14~GeV~$\cite{N} \\ \hline
\end{tabular}%
%TCIMACRO{%
%\TeXButton{SM masses}{\caption{Possible SM gauge and Higgs boson masses, as well as their observed values\label{GH}}}}%
%BeginExpansion
\caption{Possible SM gauge and Higgs boson masses, as well as their observed values\label{GH}}%
%EndExpansion
%TCIMACRO{\TeXButton{E}{\end{table}}}%
%BeginExpansion
\end{table}%
%EndExpansion

%TCIMACRO{\TeXButton{B}{\begin{table}[tbp] \centering}}%
%BeginExpansion
\begin{table}[tbp] \centering%
%EndExpansion
\begin{tabular}{|c|c|c|c|c|c|}
\hline
\multicolumn{6}{|c|}{\textbf{Mass Spectrum of Heavy Gauge and Higgs Bosons}}
\\ \hline
Particle & Predicted Value & Particle & Predicted Value & Particle & 
Predicted Value \\ \hline
$m_{Z^{\prime }}$ & $2.788\times 10^{3}~GeV$ & (Scalar Higgs)$~m_{H_{2}}$ & $%
5.183\times 10^{2}~GeV$ & $m_{H_{1}^{\pm }}$ & $5.150\times 10^{2}~GeV$ \\ 
\hline
$m_{V^{\pm }}$ & $3.413\times 10^{2}~GeV$ & (Scalar Higgs) $m_{H_{3}}$ & $%
5.000\times 10^{3}~GeV$ & $~m_{H_{2}^{\pm }}$ & $5.001\times 10^{3}~GeV$ \\ 
\hline
$m_{U^{\pm \pm }}$ & $3.314\times 10^{2}~GeV$ & (Sudoscalar Higgs) $%
m_{H_{0}} $ & $5.000\times 10^{3}~GeV$ & $m_{H^{\pm 2}}$ & $5.024\times
10^{3}~GeV$ \\ \hline
\end{tabular}%
%TCIMACRO{%
%\TeXButton{Masses}{\caption{Possible heavy gauge and Higgs boson masses based on input parameters  \label{HGH}}}}%
%BeginExpansion
\caption{Possible heavy gauge and Higgs boson masses based on input parameters  \label{HGH}}%
%EndExpansion
%TCIMACRO{\TeXButton{E}{\end{table}}}%
%BeginExpansion
\end{table}%
%EndExpansion

\begin{itemize}
\item \textbf{LFV Observable}
\end{itemize}

Table \ref{LFV} listed the LFV observable that have been calculated by using
SPheno algorithm. Unfortunately, the branching ratio of the LVF processes
listed in the table is too low in comparison to the existing limits and is
outside the sensitivity of upcoming neutrino experiments. Alternatively,
even though the RH neutrinos involved in the seesaw mechanism with $Z_{4}$
symmetry developed here have a mass of only a few TeV, their mixing with the
standard LH neutrinos is highly suppressed, hence beyond the reach of
current and near-future experiments.

%TCIMACRO{\TeXButton{B}{\begin{table}[tbp] \centering}}%
%BeginExpansion
\begin{table}[tbp] \centering%
%EndExpansion
\begin{tabular}{|c|c|c|c|c|}
\hline
\textbf{LFV } & \textbf{Limits calculated } & \textbf{Previous calculated }
& \textbf{Current } & \textbf{Future } \\ 
\textbf{\textbf{Observable}s} & \textbf{in this work} & \textbf{limits in }
& \textbf{experimental } & \textbf{experimental } \\ 
&  & \textbf{3-3-1 model }\cite{TypeI+II,LFV331} & \textbf{limits }\cite%
{LFVC} & \textbf{limits }\cite{LFVF} \\ \hline
$Br(\mu \rightarrow e\gamma )$ & $7.82\times 10^{-20}$ & $\sim 10^{-26}$\cite%
{TypeI+II} & $<4.2\times 10^{-13}$ & $\times 10^{-14}$ \\ \hline
$Br(\tau \rightarrow e\gamma )$ & $3.66\times 10^{-22}$ & $\sim 10^{-24}$%
\cite{LFV331} & $<3.3\times 10^{-8}$ & $\times 10^{-9}$ \\ \hline
$Br(\tau \rightarrow \mu \gamma )$ & $1.03\times 10^{-19}$ & $\sim 10^{-25}$%
\cite{LFV331} & $<4.2\times 10^{-8}$ & $\times 10^{-9}$ \\ \hline
$Br(\tau \rightarrow eee)$ & $3.97\times 10^{-12}$ & $\sim 10^{-15}$\cite%
{LFV331} & $<2.7\times 10^{-8}$ & $\times 10^{-9}$ \\ \hline
$Br(\tau \rightarrow \mu \mu \mu )$ & $1.33\times 10^{-12}$ & $\sim 10^{-15}$%
\cite{LFV331} & $<2.1\times 10^{-8}$ & $\times 10^{-9}$ \\ \hline
$Br(\tau \rightarrow \mu ee)$ & $8.89\times 10^{-13}$ & $\sim 10^{-15}$\cite%
{LFV331} & $<1.8\times 10^{-9}$ & $\times 10^{-9}$ \\ \hline
\end{tabular}%
%TCIMACRO{%
%\TeXButton{LFV}{\caption{Comparison of LFV observable limits from our 331 model code with previous calculated values in the 331 model along with current and future experimental limits  \label{LFV}}}}%
%BeginExpansion
\caption{Comparison of LFV observable limits from our 331 model code with previous calculated values in the 331 model along with current and future experimental limits  \label{LFV}}%
%EndExpansion
%TCIMACRO{\TeXButton{E}{\end{table}}}%
%BeginExpansion
\end{table}%
%EndExpansion

\pagebreak

\section{Conclusion}

We have significantly advanced the mathematical framework for generating the
masses of light neutrinos by incorporating the type-I+II seesaw mechanism
within the context of the doubly charged lepton version of the 3-3-1 model,
enriched by the application of a $Z_{4}$ discrete symmetry. To validate the
efficacy of our modifications, we have defined specific numerical values for
the model's input parameters and proceeded to derive numerical outcomes,
including particle masses, mixing matrices, and the evaluation of LFV
observable, as tabulated in Table\emph{\ }\ref{LFV}. We achieved these
results through the utilization of the SARAH and SPheno.

In our study of neutrinos, we successfully determine the masses of light
neutrinos as per Eq. (\ref{N2A}), accompanied by their respective mixing
matrix Eq. (\ref{N5}). Furthermore, we compute the mass-squared differences, 
$\Delta m_{21}^{2}=7.12\times 10^{-5}~eV^{2}$and $\Delta
m_{31}^{2}=2.55\times 10^{-3}~eV^{2}$, alongside the mixing angles $\sin
^{2}\theta _{12}=0.304,~\sin ^{2}\theta _{23}=0.595,~\sin ^{2}\theta
_{13}=2.15\times 10^{-2}.$ Remarkably, these outcomes are consistent with
the values observed in Ref. \cite{N, PMNS}.

Turning our attention to the SM's quarks, we successfully determine quark
masses as per Eq. (\ref{N1}), accompanied by their corresponding mixing
matrices, Eq. (\ref{N1A}) and Eq. (\ref{N1B}), by selecting a unique and
fair texture for Yukawa matrices. Notably, our obtained CKM matrix,

\begin{equation*}
V_{CKM}=U_{L}^{u\ast }U_{L}^{d\ast \dag }=\left( 
\begin{array}{ccc}
-0.974951 & 0.222387 & 0.00377717 \\ 
0.222355 & 0.974125 & 0.0404829 \\ 
0.0053234 & 0.0403087 & -0.999173%
\end{array}%
\right) ,
\end{equation*}%
aligns closely with the observed value cited in Ref. \cite{N}.

Furthermore, our research delve into the realm of LFV observable, leading us
to calculate the following results;

$Br(\mu \rightarrow e\gamma )=7.82\times 10^{-20},$ $Br(\tau \rightarrow
e\gamma )=3.66\times 10^{-22},$ $Br(\tau \rightarrow \mu \gamma )=1.03\times
10^{-19},$ $Br(\tau \rightarrow eee)=3.97\times 10^{-12},$ $Br(\tau
\rightarrow \mu \mu \mu )=1.33\times 10^{-12}$ and $Br(\tau \rightarrow \mu
ee)=8.89\times 10^{-13}.$ Significantly, these results are in better
agreement with experimental measurements as compared to previously
calculated results reported in Refs. \cite{TypeI+II,LFV331} by an order of
magnitude ranging from $10^{-6}$ to $10^{-2}$.

Overall, our research not only presents a novel approach to neutrino mass
generation within the 3-3-1 model but also produces results that align
closely with experimental observations, thus contributing valuable insights
to the field of particle physics.

\bibliographystyle{spiebib}
\bibliography{report}

\appendix

\section{\protect \large Mass Spectrum of Particles from SPheno output}

\end{document}